\documentclass[10pt,journal,compsoc]{IEEEtran}

\usepackage[T1]{fontenc}
\usepackage{mathptmx}

\ifCLASSOPTIONcompsoc
  \usepackage[nocompress]{cite}
\else
  \usepackage{cite}
\fi
\ifCLASSINFOpdf
  \usepackage[pdftex]{graphicx}
  \usepackage{mathtools}
  \usepackage{booktabs}
    \usepackage{multirow}
    \usepackage{siunitx}

\else
\fi
\usepackage{amsmath}
\usepackage{lscape}
\usepackage{amsmath}
\begin{document}
%
\title{Dimensionality Reduction for Sentiment Classification: Evolving for the Most Prominent and Separable Features}
%
%
%
%
\author{
        Aftab Anjum,
        Mazharul Islam,
        and Lin Wang~\IEEEmembership{Member,~IEEE}
\IEEEcompsocitemizethanks{\IEEEcompsocthanksitem Aftab Anjum, Mazharul Islam, and Lin Wang are with Shandong Provincial Key Laboratory of Network Based Intelligent Computing, University of Jinan, Jinan, 250022, Shandong, China.
\protect

\IEEEcompsocthanksitem 
E-mail:{aftabanjum4451@hotmail.com} (Aftab anjum);
\newline{mazharul.cse34@gmail.com} (Mazharul Islam); {wangplanet@gmail.com} (Lin Wang) 
\protect
\IEEEcompsocthanksitem 
\# Aftab Anjum and Mazharul Islam contributed equally to this work. 
\protect

\IEEEcompsocthanksitem 
* Corresponding author: Lin Wang 
\protect\\

}
}

%
%

\markboth{IEEE TRANSACTIONS ON JOURNAL NAME,  MANUSCRIPT ID}%
{Shell \MakeLowercase{\textit{et al.}}: Bare Demo of IEEEtran.cls for Computer Society Journals}
%



\IEEEtitleabstractindextext{%
\begin{abstract}
In sentiment classification, the enormous amount of textual data, its immense dimensionality and inherent noise make it extremely difficult for machine learning classifiers to extract high-level and complex abstractions. In order to make the data less sparse and more statistically significant, the dimensionality reduction techniques are needed. But in the existing dimensionality reduction techniques, number of components needs to be set manually which results in loss of the most prominent features, thus reducing the performance of the classifiers. Our prior work, i.e., Term Presence Count (TPC) and Term Presence Ratio (TPR) have proven to be effective techniques as they reject the less separable features. However, the most prominent and separable features might still get removed from initial feature set despite having higher distributions among positive and negative tagged documents. To overcome this problem, we have proposed a new framework that consists of two dimensionality reduction techniques i.e., Sentiment Term Presence Count (SentiTPC) and Sentiment Term Presence Ratio (SentiTPR). These techniques reject the features by considering term presence difference for SentiTPC and ratio of the distribution distinction for SentiTPR. Additionally, these methods also analyze the total distribution information.  Extensive experimental results exhibit that the proposed framework reduces the feature dimension by a large scale, and thus significantly improve the classification performance.
\end{abstract}

}
\maketitle
\footnote{}
\IEEEdisplaynontitleabstractindextext
%
\IEEEpeerreviewmaketitle
\IEEEraisesectionheading{\section{Introduction}\label{sec:introduction}}
%
%
%
%

 

\IEEEPARstart{S}{entiment} is a perception, view, feeling, or appraisal of an individual for specific items, organizations, or recommendation sites whereas sentiment analysis \cite{Yu:2013:UCE:2438098.2438152} distinguishes positive and negative conclusions from the content information.

In preceding years, there is a pack of examples of overcoming adversity of sentiment analysis in various areas \cite{Zhang2017}, \cite{6487473}. These days, bunch of individuals express their assessments or sentiments on multiple blogs or forums about specific items or organizations. With the progression of time, the volume of the information is considerably expanding \cite{5380788}, and the organizations need intelligent systems to understand the positive and negative aspects of reviews to comprehend clients’ supposition.
\begin{figure}
    \centering
    \includegraphics[width=.485\textwidth]{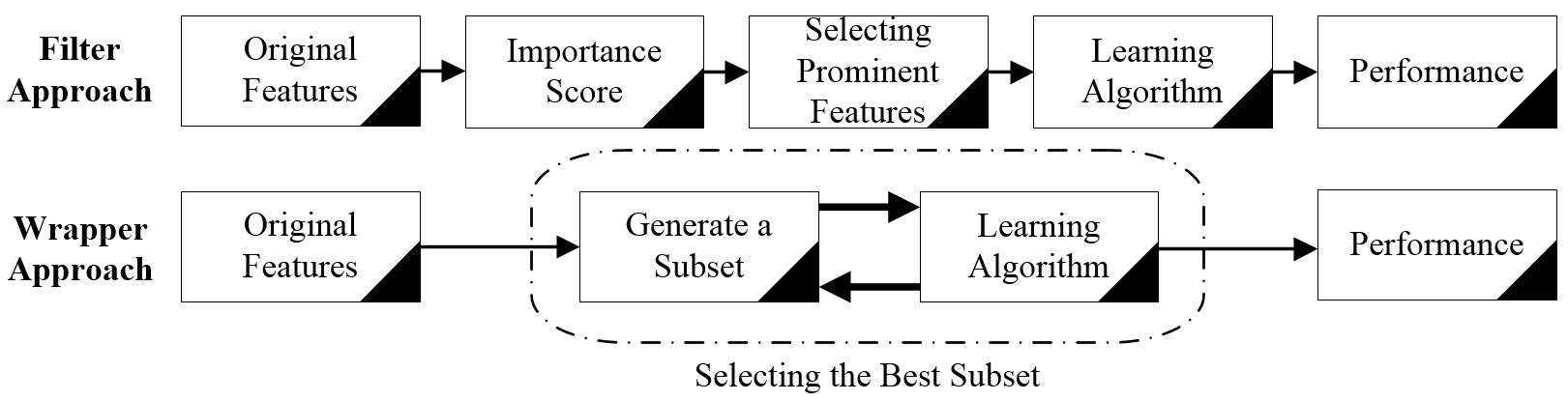}
    \caption{Shows workflow of different feature reduction approaches. Depending on user specified values, filter-based methods select the subset of features. Contrarily, wrapper-based approaches follow a greedy search algorithm that intends to create an optimal feature subset iteratively based on the classifier performance. Filter based methods are more convenient than wrapper methods for high dimensional datasets because of their faster and forthright computation abilities.}
    \label{fig:filter_wrapper}
\end{figure}
In sentiment analysis, feature selection \cite{Cai2018FeatureSI} is a recognized approach that selects the most prominent features from the initial feature vector. Feature extraction and feature reduction \cite{feacture_reduction} are two main sub parts of feature selection. The primary purpose of feature reduction is to minimize redundancy and maximize relevance to the target levels while removing irrelevant features by selecting a subgroup of features. Feature reduction is crucial for many reasons. Some of the machine learning classifiers, for instance: Naive Bayes, are expensive to train \cite{naive_bayes}. Besides, a probabilistic based method can find the probability of the real class more conveniently if the input feature vector contains the most separable features with higher distributions. Moreover, non-probabilistic methods, such as Support Vector Machine (SVM), create hyper-plane \cite{mullen-collier-2004-sentiment} in the feature space to separate different class points. Thus, the most prominent and separable features help SVM to draw the hyper-plane more smoothly. Correspondingly feature reduction often boosts the classification accuracy by efficient model while removing the trivial features \cite{Abbasi:2008:SAM:1361684.1361685}.
Improving the accuracy and efficiency of classification methods, the reduction of features is one of the vital factors. In Fig. \ref{fig:filter_wrapper}, two main approaches for feature reduction are presented. 

In the wrapper approach \cite{wrape_method}, the feature choice procedure depends on the calculation of a particular machine learning classifier that we are attempting to fit on a given dataset. Nevertheless, the wrapper based method is slower than the filter based approach because it searches each attribute of the feature set multiple times. Common methods under wrapper based techniques are Forward Selection \cite{for_reduction}, Backward Elimination \cite{backW_elimination} and Randomized Hill Climbing \cite{hillclmbing}. Filter based methods \cite{fil_approch} use some mathematical evaluations to filter out the irrelevant features. Based on some general characteristics such as correlation, mutual information gain or distance to the class attributes, filter methods select the significant features from initial feature set. For high dimensional datasets, filter based methods are more suitable than wrapper based methods because of faster and more straightforward computation. Thus our proposed methods follow the filter based approach. The most conventional filter based approaches are Gini Index (GI) \cite{chi_square}, Information (MI) \cite{mutual_info}, Chi-square Test (X2) \cite{chi_square} and Information Gain (IG) \cite{infor_gain}.

It is not reasonable to explore every single feature in the features vector. Distinguishing features and choosing the best of those features is the most ambitious thing to attempt particularly in high dimensional data. Besides, a high computation power is required to handle this task. With the expanding dimensions, the features increase likewise. This means the sparsity of the features becomes more high dimensional. Furthermore, there could be a correlation between different dimensions, thus leading to the difficulty of defining the most relevant or prominent features. The training data leads to more sparseness which is known as the curse of dimensionality. Spareness enlarging with more features which becomes extra challenging for the classifiers to generate clear decision boundaries.We need superior methods to manage high dimensional information so we can rapidly extricate examples and bits of knowledge from it. 

\begin{figure}
    \centering
    \includegraphics[width=.485\textwidth]{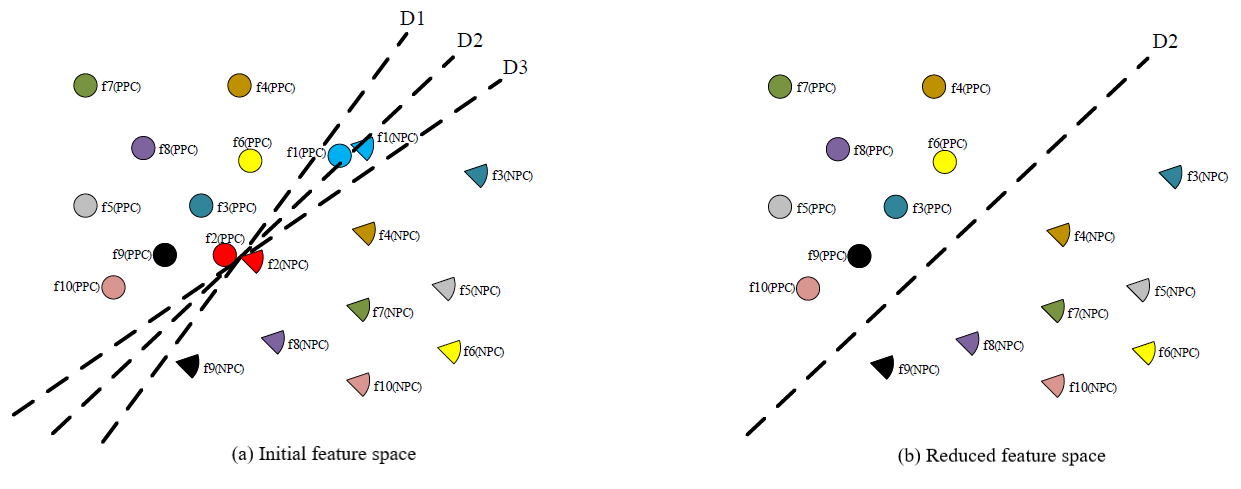}
    \caption{(a) Displays the initial feature space. (b) Shows the reduced feature space of our prior work. Circle shape represents Positive Presence Count (PPC) and cone shape represents Negative Presence Count (NPC) while various color exhibits different features.}
    \label{fig:problem_tpc_tpr}
\end{figure}

In our prior work \cite{ssci2019dr}, we proposed two dimensionality reduction methods i.e., Term Presence Ratio (TPR) and Term Presence Count (TPC) to delete those features which have nearly the same distributions among positively and negatively tagged documents. In Fig. \ref{fig:problem_tpc_tpr}, feature f1 and f2 have nearly the same distributions (i.e., Positive Presence Count (PPC) and Negative Presence Count (NPC)), which resulting in intricacy for the machine learning classifiers in order to produce precise decision boundaries. By using previously proposed TPC and TPR, we rejected these features which help classifiers to produce a precise decision boundary more conveniently. 

Regardless of deleting those less discriminative features, TPC and TPR could remove some of the important features because these two techniques delete features based on only the term presence or ratio difference, without considering the total distribution information. For instance, between feature f1 and f2, feature f2 has higher distribution but this feature will be deleted by previously proposed TPC and TPR because of having nearly same distribution. This leads to loosing important features from the initial feature set. Furthermore, the parameter value “k” was fixed by trial and error. In addition to that, we outlined a hypothesis, that our proposed methods were language independent because they were solely based on features term occurrence and did not concern about their polarity score. Though there were no such experiments to prove the hypothesis.

In our present work, we have proposed Sentiment Term Presence Count (SentiTPC) and Sentiment Term Presence Ratio (SentiTPR) that not only select or reject features based on the distribution distinction for SentiTPC and ratio of distribution distinction for SentiTPR. Both techniques also consider the total distribution information of every feature among positively and negatively tagged documents. In this manner, the loss of important features can be reduced by keeping the more prominent and most separable features. 
SentiTPC and SentiTPR when used with linear classifiers such as Support Vector Machine, help them to make decision boundary by drawing a clear gap between the feature points by deleting the less discriminative features.
Also, the probabilistic classifiers such as Random forest, Naive bayes and Logistic regression measure the probability of the class of a given sample more conveniently. This will lead the classifiers to produce better classification accuracy. Meanwhile, we have used the evolutionary algorithm to evolve the value of the parameters. Moreover, in order to prove that our proposed methods are language independent, we have used Arabic and Urdu datasets. The extensive contributions of this work are given as follows. 

\begin{itemize}
    \item We proposed two dimensionality reduction techniques (i.e., SentiTPC and SentiTPR) for sentiment classification that select or reject the features based on the distribution distinction for SentiTPC and ratio of distribution distinction for SentiTPR while considering their total distribution information.
    \item We have used evolutionary algorithm i.e., Differential evolution to evolve the hyper parameters (i.e. k and lambda) of our proposed methods.
    \item Our proposed methods are language independent. To prove this hypothesis, we used Arabic and Urdu datasets for sentiment classification.
    \item We applied the proposed techniques on different review domain datasets to show the effectiveness in multi-domain reviews. 
\end{itemize}{}
The rest of this paper is divided into 5 sections. In Section 2, we have briefly reviewed several recent representative related works. In section 3, we have introduced the general framework of this work as well as our proposed reduction techniques i.e., SentiTPC and SentiTPR. Section 4 is about the parameters analysis. Section 5 reports on the experimental results and Section 6, concludes this paper.

\section{prior work}
Over the previous years, a substantial unit of work has been accomplished in the sentiment analysis field by many researchers. Sentiment analysis may also be devised as a sentiment classification task \cite{Medhat2014SentimentAA}. In this section, prior work done on sentiment classification has been studied and summarized briefly. 
\subsection{N-gram Features for Sentiment Classification}

In sentiment classification, n-grams \cite{pang-etal-2002-thumbs} is the most popular class of features. Different n-gram approaches have obtained state-of-the-art results \cite{Abbasi:2008:SAM:1361684.1361685}, \cite{ng-etal-2006-examining}. The adoption of feature reduction approaches is required to extract the best feature set for n-gram methods. 

Fixed and variable n-grams are two subcategories of the N-gram approach. A fixed n-grams approach matches the sequence either on token or character level. In contrast, variable n-grams operate with more sophisticated linguistic phenomena. 

Bag-of words (BOWs) are typically used for text representation. Higher word n-grams (e.g., bigrams, trigrams) are the parts of the word n-grams model \cite{pang-etal-2002-thumbs}. A vector of independence tokens represents a review document. Later, the classifier model is trained by different machine learning classifiers.

Mostly unigrams, bigrams and trigrams \cite{Abbasi:2008:SAM:1361684.1361685}, \cite{ng-etal-2006-examining} are used as the N-gram approaches. Besides, four-gram methods have also been employed \cite{Argamon2007StylisticTC} for sentiment classification. Basically word n-grams model construct a feature set \cite{Argamon2007StylisticTC}, \cite{ng-etal-2006-examining}. Furthermore, much of the research work is aimed at enhancing BOW model by combining BOW with linguistic knowledge \cite{gamon-2004-sentiment}, \cite{Na2004EffectivenessOS}, \cite{ng-etal-2006-examining}, \cite{pang-etal-2002-thumbs}, \cite{Xia:2011:EFS:1924642.1924696}.

\subsection{Sentiment Classification by Employing Sentiment Dictionary}
Sentiment analysis which utilizes the sentiment dictionary is an unsupervised classification approach. Basically, the extracted tokens are assigned with some weight from the sentiment dictionary. Finally, by combining the results of the sentiment of a sentence, the class is determined. By using sentiment dictionary many researchers perform sentiment classification.

Turney and Littman, \cite{Turney:2003:MPC:944012.944013} developed an approach that can infer the sentiment arrangement of a word. For evaluating the semantic correlation between the word and the group of different paradigm words, point mutual information (PMI) and latent semantic analysis (LSA) are used. Based on the average semantic association, class of that word was classified to either positive or negative. 

By using detailed rules, Taboada et al. \cite{Taboada:2011:LMS:2000517.2000518} computed the sentiment scores of sentences, phrases, words and documents. Thereafter, the threshold setting approach was adopted by authors to find the sentiment class of the review.

Furthermore, Farhan et al. \cite{Khan2016SentiMIIP} proposed Sentiment Mutual Information (SentiMI) which is a supervised sentiment classification approach. SentiMI extracts the words through POS tagging from SentiWordNet. In sentiment dictionary, synsets represent the synonym association of the words. Finally, mutual information is estimated with considering a different class. They achieved 84\% classification accuracy for the Cornell movie review dataset.

Wu et al. \cite{wu_jiang}, extracted the features from financial data by the Apriori algorithm. Then, they obtained financial sentiment dictionary and identified sentiment tendencies by using semantic rules.

\subsection{Sentiment Classification Based on Machine Learning Classifier}

For sentiment classification, machine learning methods have been widely used at numerous levels, e.g., from the phrase or word level, to the sentence and similarly to document level.

In the earlier stages of sentiment classification, Pang et al. \cite{pang-etal-2002-thumbs} used machine learning algorithms, that are considered as a pioneer work. They used SVM, Naïve Bayes, and maximum entropy to investigate the sentiment of the movie reviews. Their empirical results manifest that SVM outperformed the other methods based on classification performance.\\
Kiritchenko et al. \cite{Kiritchenko:2014:SAS:2693068.2693087} proposed an approach that used different semantic features for sentiment classification. These semantic features were generated from the high-converge sentiment dictionary which itself was extracted from the tweets.\\
Wang et al. \cite{wang2018short} projected a method which depended upon SVM to analyse the sentiment of short text. Furthermore, the proposed method was compared with Recursive Auto Encoder and Doc2vec. Experimental verification had shown that the proposed method was more adequate for finding the sentiment of short text.

\subsection{Feature Reduction for Sentiment Classification}
In spite of the advantages \cite{Li:2006:FW:1121949.1121951} of features reduction techniques, the prior works inserted less importance on improvement. Feature reduction techniques can enhance the classification accuracy \cite{hall} by selecting the key features that can help the classifier to predict the class conveniently. There are two different types of feature reduction methods \cite{Guyon:2003:IVF:944919.944968}, i.e., univariate and multivariate. 

Chi-squared, Information gain, occurrence frequency, and log-likelihood \cite{Forman:2003:EES:944919.944974} are examples of univariate methods that study the attributes separately. Previously many researchers adopted univariate feature reduction approaches such as log-likelihood ratio and minimum frequency threshold \cite{gamon-2004-sentiment}, \cite{ng-etal-2006-examining} for sentiment classification. 

For sentiment classification \cite{Abbasi:2008:SAM:1361684.1361685}, information gain \cite{Shannon:2001:MTC:584091.584093}, has shown a great promise for text categorization. Tsutsumi et al. \cite{tsutsumi-etal-2007-movie} proposed a method that selects the features for sentiment classification using the Chi-Squired test. Even though the univariate methods are computationally faster but can be proven adverse while analyzing the distinctive features because of their inability to consider important features from the feature set. Contrarily multivariate methods choose a subset of features and evaluate based on a target classifier where predictive power is the evaluation criteria \cite{Guyon:2003:IVF:944919.944968}. Examples of multivariate methods are recursive feature elimination and decision tree models. Instead of considering the attributes separately, multivariate methods evaluate based on a group of attributes with the possible interactions. Therefore, multivariate techniques need more computational power when compared with the univariate methods.

Furthermore, some hybrid methods are also used for sentiment classification tasks. Mostly, principal component analysis (PCA) is used in dimensionality reduction for sentiment classification problems \cite{Abbasi:2008:WSA:1344411.1344413}. Recently, many reduction techniques have been applied in this field, such as geometric mean, discriminative locality alignment and harmonic mean  \cite{4760987}, \cite{4479477}. Lei Xu \cite{7430292} proposed a novel membership function to reduce the dimension of the feature set. Besides, they adopted Information gain (IG) and Singular value decomposition (SVD) for keeping the top-k values.

\subsection{Limitation of the Prior Works}
Since Bag of words (BOWs) models have high dimensional feature vector, hence they are not very useful. Feature reduction methods, however, are more effective for finding the best feature subset in order to provide improved classification accuracy. 

Although sentiment dictionary is advantageous in saving us from manually labelling the samples one by one, yet its biggest disadvantage is that its classification performance is highly dependent on the dictionaries. Currently, nearly all the sentiment dictionaries contain not enough sentiment words. 

Also, sentiment classification using machine learning classifier requires a prominent feature set to train the model. Otherwise, the classification performance becomes degraded. 

Feature reduction techniques can solve the limitations mentioned earlier,however we need simple and powerful reduction techniques to remove the unnecessary and noisy features from the initial feature set. The prominent feature set that contains highly distributed and most separable features may help the machine learning classifiers to achieve better classification performance. 
\section{Methodology}
In this section, first, the general framework of our work is given. The next subsection contains a detailed description of the pre-processing steps that are used to process the input data. Our proposed dimensionality reduction techniques, i.e., Sentiment Term Presence Ratio (SentiTPR) and Sentiment Term Presence Count (SentiTPC) are explained in the next subsections. In the final subsection, we have described the evolutionary algorithm which evolves the hyper parameters (i.e., k and lambda) that have been used in our proposed techniques.

\subsection{General Framework}
\begin{figure}
    \centering
    \includegraphics[width=.489\textwidth]{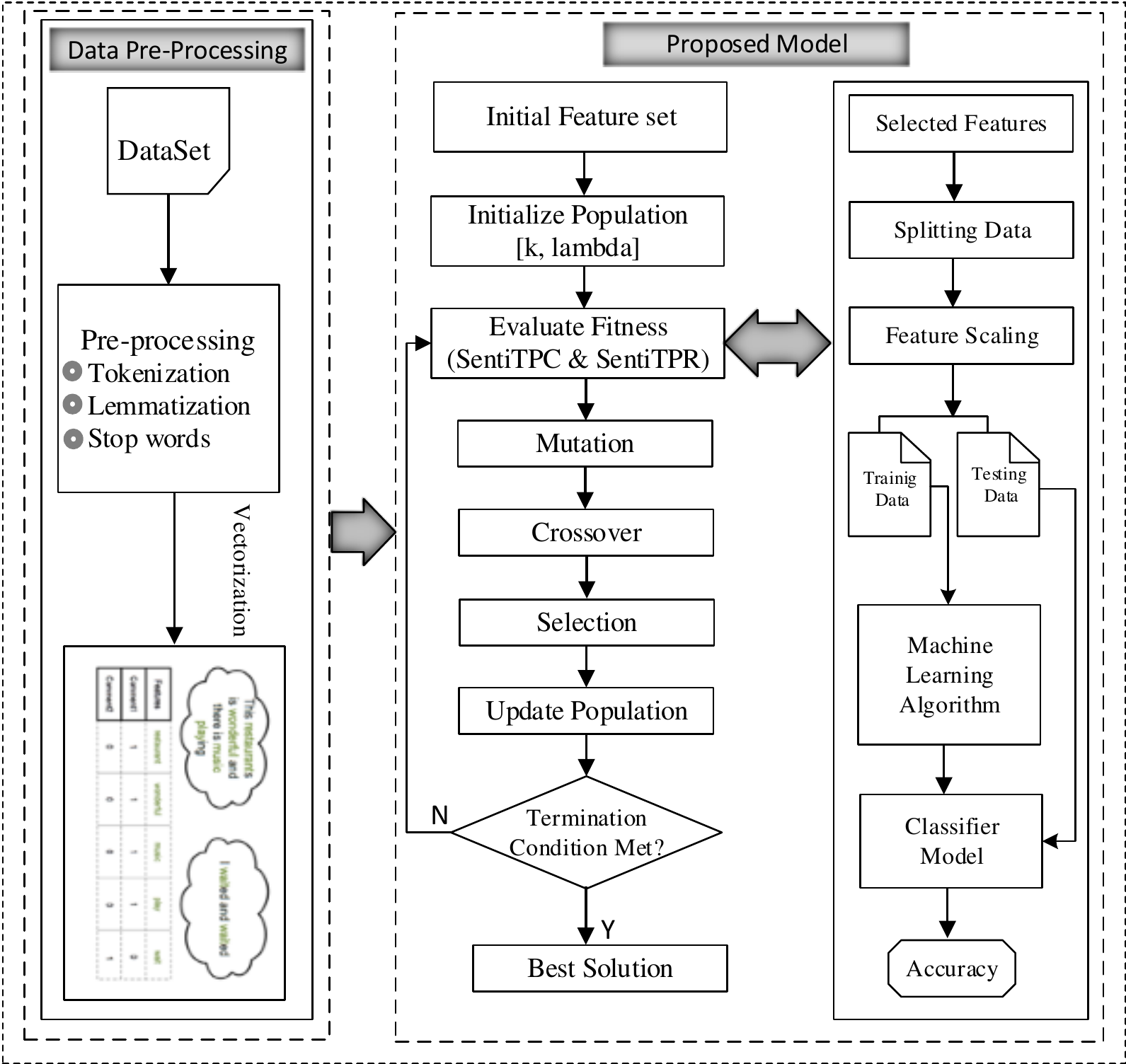}
    \caption{Explains the general framework of learning and classification process. On the left hand side in the figure, black dashed rectangle represents the Data Pre-Processing part which converts the text dataset into numerical form in order to perform numerical calculation. Whereas, on the right hand side, black dashed rectangle represents the Proposed Model. The Proposed Model starts by initializing the population (i.e., k and lambda values) in evolution process. Based on k and lambda values, SentiTPC and SentiTPR methods select the most prominent and separable features from the initial feature set and compute the accuracy value as fitness value of each individual. Then, the evolutionary process starts to evolve the population by using different genetic operators in order to maximize the accuracy value.}
    \label{fig:general_framework}
\end{figure}

The elaborated workflow of the proposed methods is shown in Fig. \ref{fig:general_framework}. The general framework is divided into two subparts, i.e., data pre-possessing and proposed model. 
Different data pre-processing techniques (i.e., tokenization, lemmatization, and stop word removal) have been applied to the entire dataset, initially. Afterwards, vectorization is used to convert all the tokens into an initial feature set. This initial feature set contains all the features present in the reviews. 

In the proposed model part as shown on the right hand side of the Fig. \ref{fig:general_framework}, first we initialize the initial population (i.e., k and lambda). After that, the proposed methods (SentiTPC and SentiTPR), based on k and lambda values of each individual, select or reject the number of features from the initial feature set. After rejecting the features from the initial feature set, the remaining features are denoted as selected feature set. After that, we adopted feature scaling, which helps to normalize the features within a fixed range. This scaling helps to improve the numeric calculations. Thereafter, training and testing set is divided from the selected feature set by 0.2 splitting ratio. That is, 20\% of the data is used for testing the machine learning model while the rest of the 80\% data is used for training. Later, different machine learning classifiers are used to calculate the accuracy value. We use the accuracy values as the fitness value of each individual in the network population. In order to maximize the fitness value, the evolutionary process starts to evolve the k and lambda values by updating the population using listed genetic operator (i.e., mutation, crossover, and selection). The whole process repeats itself until the termination condition is met.
    \subsection{Data Prepossessing}
Data pre-processing is the mechanism of cleaning and preparing the data for the classifier model. In the sentiment classification field, the dataset is mostly ambiguous, fragmentary, noisy, redundant or inconsistent, considering these are from real-world reviews given by humans. 
Any of these anomalies can downgrade the performance of the model. 
To understand the actual sentiment, these real-world data should be processed by a proper data mining approach. In this work, we have used the most common techniques to pre-process the input data which are listed below. 

        \subsubsection{Tokenization}
\begin{figure}[h]
    \centering
    \includegraphics[width=.48\textwidth]{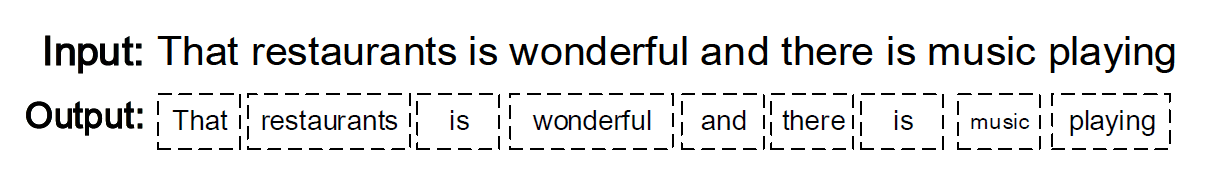}
    \caption{Tokenization process, which breaks the sentence into many tokens.}
    \label{fig:tokenization}
\end{figure}

Tokenization\cite{tokenization} is the process of splitting up a text or sentence into minimal significant units such as words, phrases, keywords, or other elements, which are known as tokens. The basic process of tokenization can be seen in Fig. \ref{fig:tokenization}.
\\

        \subsubsection{Stop Words Removal}
\begin{figure}[h]
    \centering
    \includegraphics[width=.48\textwidth]{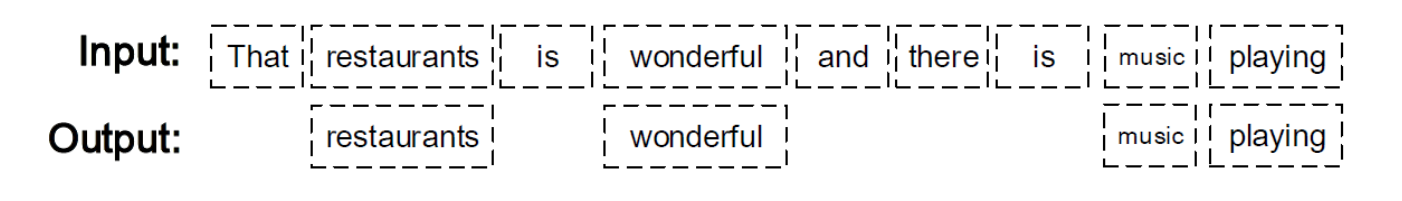}
    \caption{Stop words removal process removes the most common English words.}
    \label{fig:stop words}
\end{figure}

The most common words used in any language are known as stop words such as “the”, “is”, “an”, “all” etc. These stop words are not necessary as they do not hold any vital information. In text classification problems, the text should be classified into some predefined classes. Thus, removing the stop words from the input text would help find the actual sentiment of that particular text as it would give more concern to the rest of the words. This phenomena can be observed in Fig. \ref{fig:stop words}.

        \subsubsection{Lemmatization}
\begin{figure}[h]
    \centering
    \includegraphics[width=.48\textwidth]{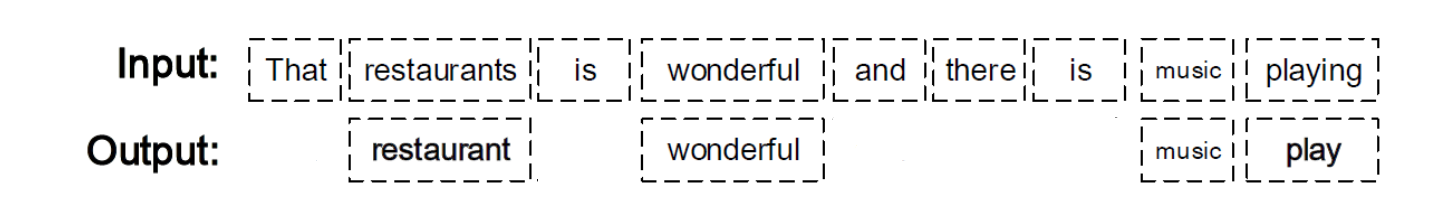}
    \caption{Lemmatization diminish a word in different forms to its root form.}
    \label{fig:Lemmatization}
\end{figure}

Any word can be represented in many forms. Lemmatization\cite{lemmatization} helps to reduce those related words to the base form. Lemmatization is very important when training the word vectors because while counting the word presence, the same word with different representations in the input text can hamper the counting of the term presence. Furthermore, it helps to produce a better feature set.The whole mechanism can be seen in Fig. \ref{fig:Lemmatization}.

        \subsubsection{Vectoraization}
\begin{figure}[h]
    \centering
    \includegraphics[width=.48\textwidth]{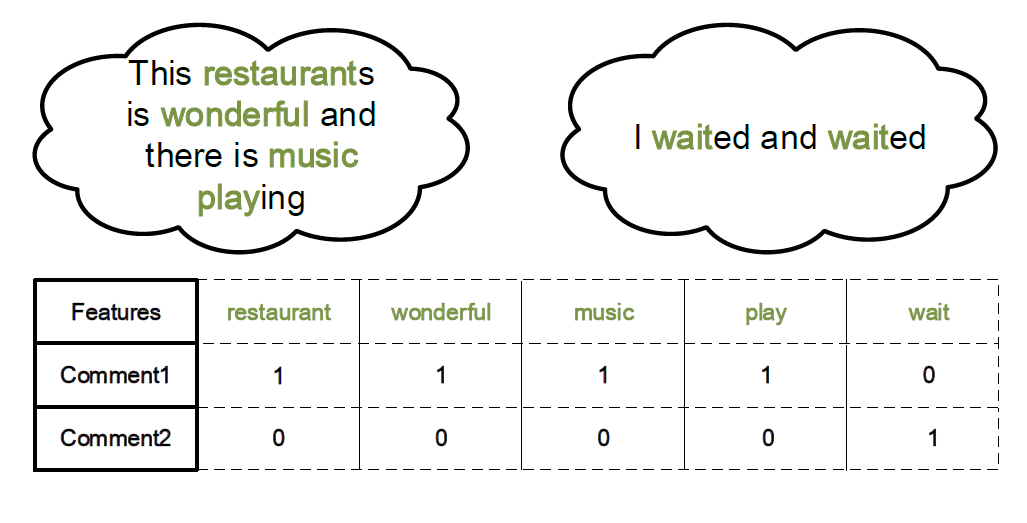}
    \caption{For every feature, vectorization approach, set 1 if the feature occurs in the review and otherwise set 0. For instance, feature "restaurant" occurs in Comment1 thus it is marked as 1 and as it does not appear in Comment2, so it is marked as 0.}
    \label{fig:vectoraization}
\end{figure}

Fig. \ref{fig:vectoraization} displays the vectorization\cite{tokenization} technique utilized in this paper. There are generally two kinds of approaches that are used to change the features into vectorized type. These approaches are: term presence and term frequency. 
Within the "term presence" approach, for every feature, if the feature occurs in review, then it will be marked as 1, otherwise it will be marked as 0. This approach is quite similar to one-hot encoding. The primary purpose is to turn the text into a vector form that we can utilize as input. Automatic text classification is a supervised machine learning task considering the classes are predefined. 
All the unique tokens are acquired after executing the pre-processing steps. With the help of vectorization, all the tokens are converted into a group of features. These corresponding tokens represent the word occurrence information in distinction to the reviews. 
During classification of movie reviews, Pang and Lee \cite{pang-etal-2002-thumbs} reported that the uni-gram features surpass bi-grams features. Moreover, another outcome of this work is that while adopting word presence rather than word frequency, this method can attain higher classification performance. Considering everything, in all our experiments, we have used term presence and uni-gram features as bag of words.

    \subsection{Sentiment Term Presence Count (SentiTPC)}
\begin{figure}
    \centering
    \includegraphics[width=.488\textwidth]{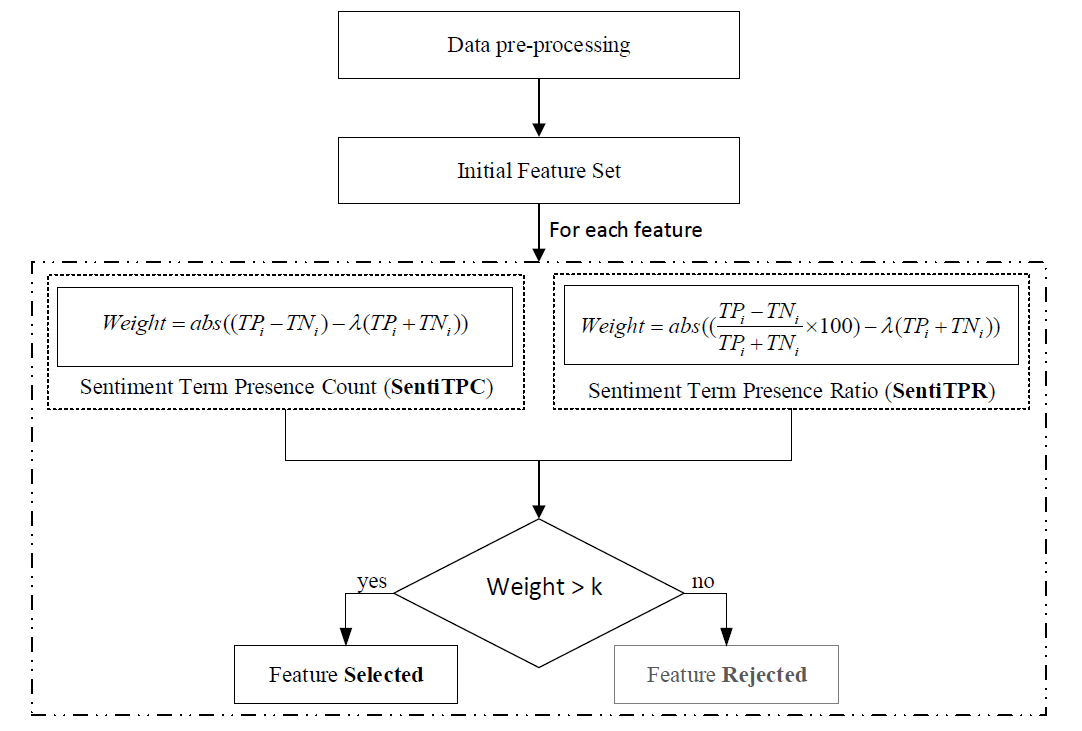}
    \caption{
    After data pre-processing, initial feature set is generated with the word presence information from vectoraization. For each feature, SentiTPC and SentiTPR use different loss functions to calculate the weight values. Afterwards, based on the evolved k value, selected feature set is composed separately while the rest of the features are removed from both methods.}
    \label{fig:sentitpc and sentitpr}
\end{figure}

The complete process of Sentiment Term Presence Count (SentiTPC) is shown in Fig. \ref{fig:sentitpc and sentitpr}. After pre-processing the input data, initial feature set has been generated, which holds the presence information of the features from the reviews.
Total presence in all of the positive reviews indicated as (TP), and contrary (TN) represents the total presence in all of the negative reviews for every features in the initial feature set.
\begin{equation}
    Distribution\,\,Distinction =  (TP_i  - TN_i)  
    \label{eq:Distribution Distinction}
\end{equation}   
Equation \eqref{eq:Distribution Distinction} exhibits the distinctness of the distribution for a particular feature. Here, i represents every feature in the initial feature set. If the distribution distinction is very small, it means the distribution is almost identical for that specific feature. In this manner, this particular feature introduces complexity for the classifier while generating the decision boundary. In our previous work \cite{ssci2019dr}, we investigated this Equation \eqref{eq:Distribution Distinction} to eliminate this kind of less distinct features, but we did not consider the total distribution information. On the assumption that the distribution distinction is very small, but particular feature has a higher number of the total distribution, this feature might be important for us. With the previous Equation \eqref{eq:Distribution Distinction}, this feature would have been eliminated. However, in our current SentiTPC method, Equation \eqref{eq:total Distinction} is proposed, which does not eliminate the above mentioned feature but rather computes the total distribution of a feature in all of the positive and negative documents. 
\begin{equation}
  Total\,\,Distribution\,\,Information = (TP_i  + TN_i )
  \label{eq:total Distinction}
\end{equation}   
To appraise the Weight of a feature the equation of SentiTPC will be,
\begin{equation}
  Weight_{TPC} = \sum\limits_{i = 1}^N {abs((TP_i  - TN_i )}  - \lambda (TP_i  + TN_i ))  
  \label{eq:EqsentiTPC}
\end{equation}   

In the above equation, N represent the total number of features and \textbf{lambda} is a parameter whose value is evolved by evolutionary algorithm. 
This weight value passes through a condition which states: if the weight value is less than the value of “k”, thereupon that distinct feature is discarded otherwise it is picked, whereas the value of “k” is settled by the evolutionary algorithm. This condition will be tested for every feature in the initial feature set. 
Using this condition those features which have nearly the same and lower distribution along positively and negatively tagged documents will be discarded. The remaining features denoted as selected features. These selected features have more distinctness and higher distribution among positively and negatively tagged documents, thus helping the classifier to make a clear decision boundary.
    \subsection{Sentiment Term Presence Ratio (SentiTPR)}
In SentiTPR, TP and TN are counted after all the pre-processing steps, including vectorization, similar to what has been done in previous SentiTPC method. SentiTPR also intends to eliminate those features which have less distinctness and distribution in all the documents represented by different classes. 
In view of this method, we have considered the ratio of the distribution distinction as well as the total distribution information of every feature in the initial feature vector. Distribution distinction ratio can be calculated by the Equation \eqref{eq: ratio of Distribution distinction} 

\begin{equation}
    Ratio\,\,of\,\,distribution \,\,distinction = (\frac{{TP_i  - TN_i }}{{TP_i  + TN_i }} \times 100)   
    \label{eq: ratio of Distribution distinction}
\end{equation} 
Total weight can be counted by the Equation \eqref{eq: eqsentiTPR} which is the combination of the Equation \eqref{eq: ratio of Distribution distinction} and Equation \eqref{eq:Distribution Distinction}.
\begin{equation}
\label{eq: eqsentiTPR}
\begin{aligned}
Weight_{TPR} &=  \sum\limits_{i = 1}^N abs((\frac{{TP_i  - TN_i }}{{TP_i  + TN_i }} \times 100) \\ 
& -\lambda (TP_i  + TN_i ))
\end{aligned}
\end{equation} 
Where the total number of features is N and \textbf{lambda} is a parameter that balances the distribution distinction ratio and total distribution. The value of \textbf{lambda} is evolved using the evolutionary algorithm. Like the previous SentiTPC method, if the weight value is less than the value of k, that feature will be deleted as it will make complexity for the classifiers. The selected features represent a reduced feature vector with higher distinctness and distribution in all of the documents that can be seen in Fig. \ref{fig:sentitpc and sentitpr}. In this manner, the classifiers will get the most prominent and separable features, which helps to make precise decision boundary and improve the classification performance.

\subsection{Evolutionary Process}
\begin{figure}[h]
    \centering
    \includegraphics[width=.489\textwidth]{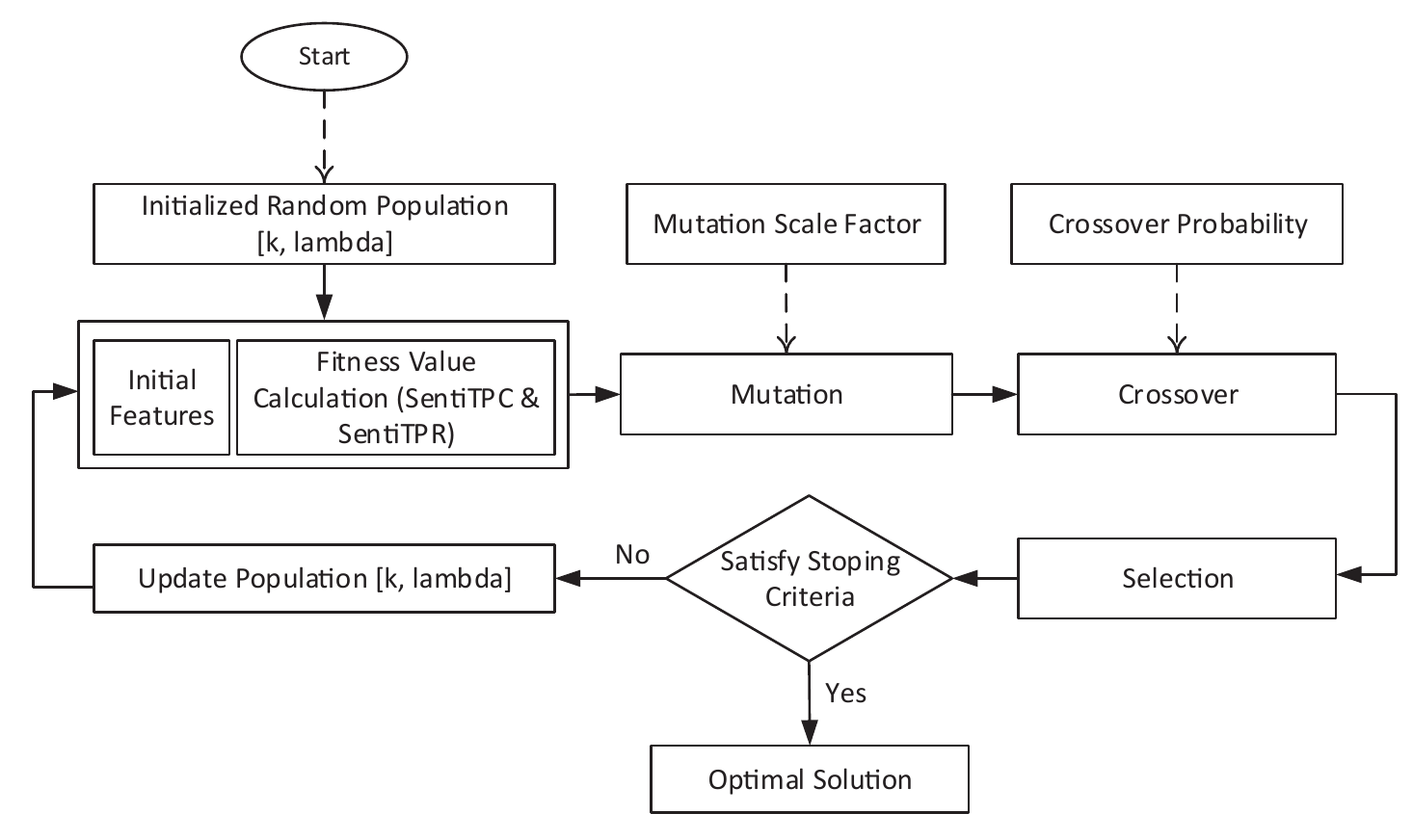}
    \caption{The basic flow of the evolutionary process is described in this figure. The process starts by initializing the population i.e., k and lambda values, then based on the k and lambda values, select the most significant features from the initial feature set and compute the fitness or accuracy value. Then, by using genetic operators to update the population and calculate the effectiveness of the updated individuals.}
    \label{fig:evolve_process}
\end{figure}

The best possible values of decision variables are obtained through a procedure called optimization which uses a given set of constraints and a selected optimization objective function to find the best possible solution. The basic flow of the evolutionary process can be seen in fig \ref{fig:evolve_process}. The first step of this evolutionary process (i.e., Differential Evolution) is to set up the necessary parameters (i.e., population size, number of generations, mutationand crossover probability), and initialize the random population (i.e., k and lambda values according to their respective bound range). After that, the proposed methods (SentiTPC and SentiTPR), based on k and lambda values of each individual, select the number of features from the initial feature vector and compute the accuracy values. We use the accuracy values as the fitness value for each individual in the network population.
In order to maximize the fitness value, the evolutionary process selects the most significant and relevant features from the initial feature set by updating the population (k and lambda values) using listed genetic operators (i.e., mutation, crossover, and selection). The process repeats itself until the termination condition is met.  
\subsubsection{Differential Evolution}
Differential Evolution\cite{DE} (DE) is a population based stochastic optimization algorithm for solving the nonlinear optimization problem. Without having prior knowledge, DE can locate useful solutions for complex problems. In order to utilize exploration and exploitation operations, Differential Evolution uses two genetic operators which are cross-over and mutation. In contrast, the selection operator is used to lead the search towards the prospective regions and to ensure that the best solution is certainty part of the next generation.
\\ 
Presently, there are several modifications of DE \cite{DEv2}, \cite{DEv3}, \cite{DEv4}. The explicit version which is used during this investigation is called the DE $"best1bin"$ scheme.\\
The optimization objective is to diminish the value of this objective function f(a),
\begin{equation}
{\rm{minimize}}(f(a))
\end{equation}
by optimizing the values of its parameters:
\begin{equation}
a = \left[ {a_1 ,a_2 ,a_3 ,a_4 ,........., a_x } \right],a \in R^x  
\end{equation} 
Where, a is a vector that consists of x objective function parameters. Every maximization problem can be turned into a minimization problem by multiplying the objective function with -1. The population matrix can be shown as
\begin{equation}
a_{p,i}^j  = \left[ {a_{p,1}^j ,a_{p,2}^j ,a_{p,3}^j ,a_{p,4}^j ,..........., a_{p,x}^j } \right]
\end{equation}
where, j is the generation, P act as the population size $p = 1,2,3,......,P $ and $i = 1,2,3,...,x$.    

\paragraph{Initialization}
To seek optimum values, population must be initialized as a starting point, and the random values from within the boundary constrains can be used to initialize the initial population $a_{p,i}$. 
\begin{equation}
a_{p,i}  = a_{p,i}^L  + rand()\times(a_{p,i}^U  - a_{p,i}^L )
\end{equation}
where, $a_{i}^L$, $a_{i}^U$ set as lower and upper bound for the variable $a_{i}$ respectively and rand() denotes a uniformly distributed random value within the range [0.0, 1.0]. 
\paragraph{Mutation}
For a given parameter vector $a_{p,i}^j$, select three other vectors $a_{{r_1}p}^j$, $a_{{r_2}p}^j$ and $a_{{r_3}p}^j$ randomly. A mutant vector $d_p^{j + 1}$ is generated by adding the weighted difference of two of the vectors to the third vector.

\begin{equation}
 d_{p}^{j + 1}  = a_{{r_3 }p}^j + F(a_{{r_2 }p}^j - a_{{r_1 }p}^j)
\end{equation} 

Whereas, r1, r2 and r3 are randomly chosen indexes. The values of these indexes are not only different from each other but also from the running index, i. The amplification of the differential variation $(a_{{r_2 }p}^j - a_{{r_1 }p}^j)$ is controlled by $F$ which is a real and constant factor whose range is given by $\in [0.5,1]$. 

\paragraph{Cross over}
In DE, cross-over operator is used in order to increase the variation in the generated parameter vectors.
 A trial vector $R_{p,i}^{j + 1}$ is evolved from the target vector ${a_{p,i}^{j + 1} }$ and mutant vector $d_{p,i}^{j + 1}$. The crossover probability, ${\rm{Rp}} \in [0,1]$ is predefined in the classic version of DE. When ${\rm{Rp}}$ is compared with a random number $rand \in [0,1]$. If the random number is less than or equal to ${\rm{Rp}}$ or ${\rm{i = k}}_{rand}$ then the trial parameter is handed down from the mutant vector $d_{p,i}^{j + 1}$. In other case, the trail parameter is imitate from the vector ${a_{p,i}^{j + 1} }$.
\begin{equation}
 \mathop {R_{p,i}^{j + 1} }\nolimits_{}^{}  = \left\{ \begin{array}{l}
 \mathop {d_{p,i}^{j + 1} }\nolimits^{} {\rm{  if\, rand() }} \le {\rm{ Rp\quad  or\quad  i = k}}_{rand}  \\ 
  \\ 
 {\rm{ }}\mathop {a_{p,i}^{j + 1} }\nolimits^{} {\rm{ if\, rand()  >  Rp\quad  or\quad  i}} \ne {\rm{k}}_{rand}  \\ 
 \end{array} \right.
\end{equation} 
Where, ${\rm{i = k}}_{rand}$ is a integer random number between [1, D] and D=1,2,3,.....,x. 

\paragraph{Selection}
The selection operator in DE, ensure that the best solution will be part of the next generation. In order to achieve the trial vector, $R_{p,i}^{j + 1}$ is compared to the target vector ${a_p^j }$ using the greedy criterion. If the target vector shows higher output cost function value than the trail vector then the trial vector $R_{p,i}^{j + 1}$ is inherited to the next generation. Otherwise target vector ${a_p^j }$ is proceed.   
\begin{equation}
 \mathop {a_p^{j + 1} }\nolimits_{}^{}  = \left\{ \begin{array}{l}
 \mathop {R_{p,i}^{j + 1} }\nolimits^{} {\rm{  if\,  }}f(R_p^{j + 1} ) < f(a_p^j ) \\ 
 {\rm{ }}\mathop {a_p^j }\nolimits^{} \quad{\rm{                  otherwise}} \\ 
 \end{array} \right.
\end{equation}

\section{Parameter Analysis}
In this section, we explained the parameters we have adopted in our proposed methods (i.e., SentiTPC and SentiTPR). In our experiments, for the proposed methods, the parameters values “k” and “lambda” was chosen by evolutionary process (i.e., Differential Evolution).       
\subsection{k value}
Based on the value of k, the proposed SentiTPC and SentiTPR determine which feature should be eliminated in distinction to the initial feature vector. For SentiTPC, the value of k evolves from the interval [1, 30], and for SentiTPR the value of k evolves from the interval [1, 50]. For the SentiTPR method, larger k values impose a higher constraint on the proposed method to remove all of the irrelevant and insignificant features from the initial feature set. On the other hand, for the SentiTPC method, much higher k values remarkably reduce the feature set size, which consequently eliminates the important and prominent features from the initial feature set and reduces the accuracy. Therefore, for the sake of maximizing the accuracy of the proposed methods, the values of k will be chosen by the evolutionary process that can produce the best feature set which can achieve the highest accuracy.
    \subsection{lambda value}
In both of the proposed methods, we compute the Weight values which help us to evaluate that whether a particular feature is prominent or not. In this work, to select a feature, we use total distribution information along with the distribution distinction for SentiTPC and total distribution information along with ratio of the distribution distinction for SentiTPR. In this way, we can discover better feature set which aids to enhance the classification performance. For calculating the Weight by SentiTPC, lambda value helps to keep the balance between distribution distinction and the total distribution information. Similarly, for SentiTPR, lambda value assists to keep the equilibrium between ratio of the distribution distinction and the total distribution information. The lambda value will also be evolved by the evolutionary process from the range [0.1, 0.5]. Thus, the best lambda value will be selected by the evolutionary process which produces the best feature set.

\section{Experimental Verification}
For experimental verification, we empirically evaluate our proposed reduction techniques on sentiment classification across 6 review datasets from different domain over 4 classification algorithm along with 6 state-of-the-art reduction techniques. Later we explain the feature reduction with SentiTPC and SentiTPR and in the end we analyze hyper-parameters sensitivity study.

\subsection{Corpora Description}
We used six datasets from different review domains to validate our proposed reduction techniques.The characteristics of these datasets can be seen in the Table \ref{table:Chars of Data Sets}. Dataset 1\cite{dataset1} and dataset 2\cite{dataset2} are movie review datasets, which were generated by Bo Pang and Lillian Lee. 
Dataset 3\cite{dataset3} was created by N. A. Abdulla and N. Mahyoub for Arabic sentiment classification in 2014. This dataset contains 1982 tweets about various topics such as politics and arts. As a binary classification dataset, the class distribution is fifty percent for each class. For positive and negative reviews, the average words in each tweet are 7.19 and 9.97 respectively.  
Dataset 4 and Dataset 5 were taken from the paper \cite{dataset4and5}. Each dataset contains one thousand sentences, labelled with either positive or negative sentiment. These datasets are extracted from consumer products and restaurants review. Dataset \cite{dataset6} is roman Urdu dataset consists of reviews from e-commerce websites, comments on public Facebook pages, and Twitter accounts. The original dataset contained twenty thousand reviews and had a corresponding sentiment attached to it, which was positive, negative, and neutral. We removed the neutral reviews because this work concentrates only on binary sentiment classification of different review domains.
We extracted unigram features from all the above mentioned datasets.
\begin{table}[h]
	\centering
	\caption{Typical attributes of the datasets.}
	\begin{tabular}{p{1cm} p{1.2cm} p{1.1cm}  p{1.1cm}  p{0.9cm} p{0.9cm}}
		\hline
		Domain & Dataset & Num. of Reviews  & Num. of Features & Positive Reviews & Negative Reviews\\    \hline
		\multirow{2}{*}{Movie} & dataset 1 & 2,000   & 41,401 & 1,000 & 1,000\\ & dataset 2 & 1,400  & 36,265 & 700 & 700\\
		Twitter & dataset 3 & 1,982  & 7,093 & 991 & 991\\
		\multirow{3}{*}{\parbox{1cm}{Product \& Service}}
		& dataset 4 & 1,000  & 1,680 & 500 & 500\\
		& dataset 5 & 1,000  & 1,501 & 500 & 500\\
		& dataset 6 & 10,572  & 27,063 & 5,286 & 5,286\\
		\hline
		\label{table:Chars of Data Sets}
	\end{tabular}
\end{table}

\subsection{Classification Techniques}

\begin{enumerate}
  \item\textbf{Support Vector Machine:}
 Support Vector Machine\cite{SVM} is widely employed to deal with text classification problems, but with the bigger dataset or overlapping classes, the performance can be downgraded because of high training time. On that note, if the irrelevant or overlapped features are removed from the feature space, then SVM can easily decide the soft boundary, thus improves the generalization ability of the model.
  \item \textbf{Logistic Regression:}
Logistic Regression\cite{LR} measures the relationship between a dependent variable such as class and one or more than one independent variable such as features by estimating their probabilities. LR works better when the cognate attributes are being eliminated from the feature vector.

  \item\textbf{Naive Bayes:}
Naïve bayes\cite{NB} is a general approach for sentiment classification, which utilizes the well-known Bayes theorem. As a probabilistic method, NB discovers the strong probability regarding the features with strong independence assumptions.
  
  \item\textbf{Random Forest:}
Random Forest\cite{RF} operates as an ensemble learning approach. From an incoherently selected subset of training data, RF forms several decision trees. RF can draw complex decision boundaries as it uses an ensemble of trees.
 
\end{enumerate}

\subsection{Reduction Techniques}   
\begin{enumerate}
  \item \textbf{Principal Component Analysis:} As a feature reduction technique, Principal Component Analysis\cite{PCA1} shrinks an extensive number of features to a small set of features yet holding the same information. However, for high dimensional data sets, PCA is quite expensive to compute.  
  \item \textbf{Chi-square:} The Chi-Square (Chi2)\cite{chi2} is a statistic that is frequently employed for checking out relationships between categorical variables. The null speculation of the Chi-Square reveals that there are no certain relationships on the categorical variables within the population; they are autonomous. 
  
   \item\textbf{Spare Random Projection:} In Sparse random projection\cite{SRP}, random matrix is commonly used to convert high dimensional data into a low dimensional subspace.
   
    \item \textbf{Gaussian Random Projection:} Gaussian Random Projection\cite{GRP} projects the original input space on a randomly generated matrix, the parts of which are derived from the given-below Equation \eqref{eq:GRP}.
     \begin{equation}
      N\left( {0,\frac{1}{{{n_{components}}}}} \right)\ 
      \label{eq:GRP}
      \end{equation} 
    
     \item \textbf{Isometric Mapping:}
     Isometric Mapping\cite{Isomap} searches for a lower-dimensional embedding to maintain geodesic distances between all points.
     
      \item\textbf{Kernel principal component analysis:} For optimal dimensionality reduction of the non-linear dataset, a kernel function is employed by kernel PCA to project the dataset into higher dimensional feature space. This higher dimensional feature space is linearly separable which is very much alike to the idea of Support Vector Machine.

\end{enumerate} 
\begin{table}[h]
	\centering
	\caption{Competitive approaches.}
	\label{table:competitive approach}
	\begin{tabular}{l}
		\hline
		Support Vector Machine (SVM) + None\\    
		Support Vector Machine (SVM) + Principle Component Analysis (PCA)\\
		Support Vector Machine (SVM) + Chi-square (Chi2)\\
		Support Vector Machine (SVM) + Spare Random Projection (SRP)\\
		Support Vector Machine (SVM) + Gaussian Random Projection (GRP)\\
		Support Vector Machine (SVM) + Isometric Mapping (ISOMAP)\\
		Support Vector Machine (SVM) + Kernel PCA (K-PCA)
		
		\\\hline
        Logistic Regression (LR) + None\\    
		Logistic Regression (LR) + Principle Component Analysis (PCA)\\
		Logistic Regression (LR) + Chi-square (Chi2)\\
		Logistic Regression (LR) + Spare Random Projection (SRP)\\
		Logistic Regression (LR) + Gaussian Random Projection (GRP)\\
		Logistic Regression (LR) + Isometric Mapping (ISOMAP)\\
		Logistic Regression (LR) + Kernel PCA (K-PCA)
		\\\hline
		Random Forest (RF) + None\\
		Random Forest (RF) + Principle Component Analysis (PCA)\\
		Random Forest (RF) + Chi-square (Chi2)\\
		Random Forest (RF) + Spare Random Projection (SRP)\\
		Random Forest (RF) + Gaussian Random Projection (GRP)\\
	    Random Forest (RF) + Isometric Mapping (ISOMAP)\\
		Random Forest (RF)) + Kernel PCA (K-PCA)
		\\\hline

		Naive Bayes (NB) + None \\
		Naive Bayes (NB) + Principle Component Analysis (PCA)\\
		Naive Bayes (NB) + Chi-square (Chi2)\\
		Naive Bayes (NB) + Spare Random Projection (SRP)\\
		Naive Bayes (NB) + Gaussian Random Projection (GRP)\\
		Naive Bayes (NB) + Isometric Mapping (ISOMAP)\\
		Naive Bayes (NB) + Kernel PCA (K-PCA)
		\\ 	\hline
	\end{tabular}
\end{table}
\subsection{Parameter Settings}
To appraise the performance of our proposed dimensionality reduction methods, we chose four typical machine learning classifiers along with six widely used dimension reduction techniques which are mentioned in the above subsections and also can be seen in Table \ref{table:competitive approach}. During the experiments, for some datasets, some reduction techniques produced null values during the reduction of feature size. Ultimately, when these reduced feature-sized datasets were put into the classifiers, classifiers produced value error. Therefore, in order to compile these datasets, we explicitly remove all the null values from them. 
As we know all the reduction techniques require number of components to reduce the feature size. Therefore, for a legitimate comparison, we put the equal number of component values among all the reduction techniques. We get values of the number of components from our proposed methods for every dataset and classifier; the details can be seen in Fig. \ref{fig:Number of features of Datasets}. 

Moreover, we used four different machine learning classifiers in our experiments as these are the most common methods in the sentiment classification field. We implement the default parameter settings for SVM classifier, i.e., value of kernel function was assigned as linear and random state as zero while all the other parameter values were set as default as listed in sklearn toolkit. For naive Bayes Classifier, we used sklearn toolkit once more, in which  we used a multinomial event model with alpha value equals 1.0 whereas class prior and fit prior were set none and true respectively. Similarly for LR and RF, we used sklearn toolkit. For both of these classifiers, the value of the random state was set to zero, whereas for RF, number of estimator value was set equal to 10 and entropy as criterion parameter.\\

Furthermore, the basic parameters setting of Evolutionary algorithm i.e., Differential Evolution is listed in Table \ref{table:Parameters setting of Differential Evolution}. During the experiments, the total number of Generation used for dataset 3, 4 and 5 are 1000, whereas for dataset 1, 2 and 6 are 100. All the other parameters of Differential Evolution are set according to default parameters as listed in the $''scipy.optimize.differential.evolution''$ library.
\begin{table}[h]
	\centering
	\caption{Parameters settings of differential evolution.}
	\begin{tabular}{p{1.7cm} p{1.7cm}  }
		\toprule
		Parameters & Values  \\ 
		\bottomrule
		Cross over & 0.7  \\ 
		Mutation &(0.5, 1.0)\\ 
		Pop Size & 15\\ 
		Workers & 1\\ 
       Strategy & best1bin\\ 
       Updating & immediate\\ 
       Seed & N/A\\ 
       Callback &  N/A\\ 
		\midrule
	\label{table:Parameters setting of Differential Evolution}
	\end{tabular}
\end{table}

\subsection{Performance Evaluations}
In this subsection, we exhibit the results and discuss the contributions of this work. We have used four machine learning classifiers, including LR, SVM, NB, and RF with our proposed SentiTPC and SentiTPR. For further comparison, we employed six additional reduction techniques i.e., PCA, Chi-square, Kernel PCA, Gaussian Random Projections, Sparse Random Projections, and Isometric Mapping.\\
From Table \ref{table:gene accuracy table}, we can notice that among all the classifiers, LR shows the highest accuracy and average F measure for our proposed methods (SentiTPC and SentiTPR), for all the datasets except for dataset 3 and dataset 6. For dataset 1, LR SentiTPR shows the highest accuracy score of 96.50\% among all other reduction and non-reduction techniques. Similarly, for dataset 2 and dataset 4, LR sentiTPR outperforms every other technique by accuracy score of 97.14\% and 81.00\% respectively. On the other hand, for dataset 5, LR sentiTPC shows the highest accuracy score of 86.50\%, whereas, for dataset 3 and dataset 6, NB sentiTPC and NB SentiTPR show the highest accuracy scores of 89.16\% and 82.97\% respectively.
In terms of NB classifier, while using all the reduction techniques except Chi2* and Chi2**, it scales the input, thereafter generates negative values. Whereas, NB utterly needs positive values for computation due to this, there is no experiment result mention in the Table \ref{table:gene accuracy table} for these reduction techniques.
\begin{table*}
 \caption{Generalization accuracy and average F-measure}
  \begin{tabular}{llSSSSSSSSSSSS}
    \toprule
    \multirow{2}{*}{Methods} &
    \multirow{2}{*}{Techniques} &
      \multicolumn{2}{c}{Dataset 1} &
      \multicolumn{2}{c}{Dataset 2} &
      \multicolumn{2}{c}{Dataset 3} &
      \multicolumn{2}{c}{Dataset 4} &
      \multicolumn{2}{c}{Dataset 5} &
      \multicolumn{2}{c}{Dataset 6} \\
     & & {ACC.} & {Avg.FM} & {ACC.} & {Avg.FM} & {ACC.} & {Avg.FM}& {ACC.} & {Avg.FM}& {ACC.} & {Avg.FM}& {ACC.} & {Avg.FM} \\
     \midrule
     SVM & None & 79.00 & 78.93  &  75.71 & 75.66 &  65.23 & 63.96  & 68.00 & 68.00 &  65.00 & 64.30 & 70.44 & 70.33\\
     & Chi2* & 53.00 & 48.16 & 52.85 & 49.98 & 74.81 & 74.03 & 66.50 & 64.35 & 63.00 & 60.07 & 59.19 & 57.60 \\
     & Chi2** & 51.75 & 38.90 & 55.35 & 54.34 & 74.55 & 74.46 & 58.50 & 53.56 & 58.00 & 53.50 & 55.13 & 49.77 \\
     & PCA* & 79.00 & 78.93 & 75.71 & 75.66 & 74.05 & 63.96 & 68.00 & 67.84 & 72.50 & 71.91 & 72.86 & 72.83 \\
     & PCA** & 79.00 & 78.93 & 75.71 & 75.66 & 65.23 & 63.96 & 68.00 & 68.00 & 65.00 & 64.30 & 70.44 & 70.33 \\
     & GRP* & 73.50 & 73.49 & 72.85 & 72.76 & 66.24 & 66.09 & 63.50 & 63.45 & 66.00 & 65.65 & 61.08 & 61.09 \\
     & GRP** & 72.25 & 72.12 & 74.28 & 74.25 & 66.49 & 65.72 & 68.50 & 68.48 & 67.50 & 67.14 & 69.97 & 69.94 \\
     & KPCA* &79.00 & 78.93 & 75.71 & 75.66 & 74.05 & 73.57 & 65.50 & 65.03 & 73.00 & 72.19 & 73.28 & 73.28 \\
     & KPCA** & 79.00 & 78.93 & 75.71 & 75.66 & 65.23& 63.96 & 68.00 & 68.00 & 65.00 & 64.30 & 70.44 & 70.33 \\
     & SRP* & 71.00 & 70.83 & 72.14 & 71.92 & 63.22 & 62.61 & 66.00 & 65.78 & 65.50 & 65.43 & 63.12 & 63.12 \\
     & SRP** & 73.50 & 73.25 & 73.92 & 73.88 & 65.23 & 64.41 & 67.50 & 67.50 & 67.50 & 67.06 & 68.41 & 68.38 \\
	 & ISOMAP* & 71.70 & 70.83 & 72.14 & 71.92 & 63.22 & 62.61 & 66.00 & 65.74 & 65.50 & 65.43 & 61.89 & 60.33 \\
     & ISOMAP** & 73.50 & 73.25 & 73.92& 73.88 & 65.23 & 64.41 & 67.50 & 67.50 & 67.50 & 67.06 & 56.17 & 54.88 \\
     & SentiTPC & 90.50 & 90.50 & 92.85 & 92.85 & 81.36 & 81.30 & 75.50 & 74.89 & 79.50 & 79.22 & 76.78 & 75.98 \\
     & SentiTPR & 96.00 & 96.00 & 95.00 & 94.99 & 77.32 & 76.28 & 74.00 & 73.78 & 75.00 & 73.84 & 73.80 & 73.32 \\
  LR & None & 79.00 & 78.93 & 75.00 & 74.93 & 82.61 & 82.57 & 69.50 & 69.33 & 71.00 & 70.64 & 74.79 & 74.79 \\
     & Chi2* & 55.25 & 48.23 & 56.07 & 54.76 & 75.56 & 74.64 & 60.00 & 58.65 & 66.00 & 63.72 & 57.35 & 55.64 \\
     & Chi2** & 51.25 & 37.15 & 54.28 & 51.54 & 74.30 & 74.00 & 72.00 & 69.96 & 60.50 & 55.47 & 53.52 & 44.64 \\
     & PCA* & 79.00 & 78.93 & 75.00 & 74.93 & 86.39 & 86.39 & 72.00 & 71.71 & 72.00 & 71.86 & 74.04 & 74.03 \\
     & PCA** & 79.00 & 78.93 & 75.00 & 74.93 & 82.61 & 82.57 & 69.50 & 69.33 & 71.00 & 70.64 & 74.70 & 74.69 \\
     & GRP* & 75.75 & 75.65 & 70.35 & 70.34 & 72.04 & 72.03 & 60.00 & 60.00 & 67.00 & 66.92 & 58.53 & 58.53 \\
     & GRP** & 76.00 & 75.99 & 72.85 & 72.28 & 78.08 & 78.04 & 70.00 & 69.85 & 70.00 & 69.56 & 70.96 & 70.97 \\
     & KPCA* & 79.00 & 78.93 & 75.00 & 74.93 & 86.39 & 86.39 & 68.50 & 68.00 & 72.00 & 71.77 & 74.79 & 74.80 \\
     & KPCA** & 79.00 & 78.93 & 75.00 & 74.93 & 82.61 & 82.57 & 69.50 & 69.33 & 71.00 & 70.64 & 74.75 & 74.74 \\
     & SRP* & 73.38 & 73.50 & 71.78 & 71.78 & 74.81 & 74.81 & 65.00 & 64.87 & 70.00 & 69.99 & 59.43 & 59.65 \\
     & SRP** & 74.93 & 74.99 & 73.57 & 73.55 & 79.59 & 79.57 & 69.50 & 69.30 & 74.50 & 74.26 & 71.96 & 71.93 \\
	 & ISOMAP* & 73.38 & 73.50 & 71.78& 71.78 & 74.81 & 74.81 & 65.00 & 64.87 & 70.00 & 69.90 & 60.61 & 60.62 \\
	 & ISOMAP** & 75.00 & 74.93 & 73.57& 73.55 & 79.59 & 79.59 & 69.50 & 69.37 & 74.50 & 74.26 & 60.37 & 60.10 \\
     & SentiTPC & 91.25 & 91.24 & 93.21 & 93.21 & 87.15 & 87.15 & 78.50 & 77.82 & \textbf{86.50} & \hspace{0.2cm}\textbf{86.37} & 78.43 & 78.21 \\
     & SentiTPR & \textbf{96.50} &\hspace{0.1cm}\textbf{ 96.49} & \textbf{97.14} & \hspace{0.1cm} \textbf{97.14} & 88.91 & 88.91 & \textbf{ 81.00} &\hspace{0.1cm} \textbf{80.62} & 82.50 & 82.12 & 81.27 & 81.09 \\
  NB & None & 79.25 & 79.23 & 76.42 & 76.42 & 82.87 & 82.66 & 70.50 & 69.86 & 75.00 & 74.91 & 78.81 & 78.82 \\
     & Chi2* & 55.00 & 46.89 & 56.07 & 45.15 & 71.78 & 71.54 & 57.50 & 56.31 & 57.50 & 55.99 & 57.25 & 55.74 \\
     & Chi2** & 50.50 & 33.89 & 51.78 & 35.33 & 74.55 & 73.82 & 67.00 & 63.21 & 59.50 & 54.66 & 56.78 & 51.01 \\
     & PCA* & N/A & N/A & N/A & N/A &N/A  & N/A & N/A & N/A & N/A & N/A & N/A & N/A \\
     & PCA** & N/A & N/A & N/A & N/A & N/A & N/A & N/A & N/A & N/A & N/A & N/A & N/A \\
     & GRP* & N/A & N/A & N/A & N/A & N/A & N/A & N/A & N/A & N/A & N/A & N/A & N/A \\
     & GRP** & N/A & N/A & N/A & N/A & N/A & N/A & N/A & N/A & N/A & N/A & N/A & N/A \\
     & KPCA* & N/A & N/A & N/A & N/A & N/A & N/A & N/A & N/A & N/A & N/A & N/A & N/A \\
     & KPCA** & N/A & N/A & N/A & N/A & N/A & N/A & N/A & N/A & N/A & N/A & N/A & N/A \\
     & SRP* & N/A & N/A & N/A & N/A & N/A & N/A & N/A & N/A & N/A & N/A & N/A & N/A \\
     & SRP** & N/A & N/A & N/A & N/A & N/A & N/A & N/A & N/A & N/A & N/A & N/A & N/A \\
	 & ISOMAP* & N/A & N/A & N/A & N/A & N/A & N/A & N/A & N/A & N/A & N/A & N/A & N/A \\
	 & ISOMAP** & N/A & N/A & N/A & N/A & N/A & N/A & N/A & N/A & N/A & N/A & N/A & N/A \\
     & SentiTPC & 87.50 & 87.50 & 87.14 & 87.14 & \textbf{ 89.16} & \hspace{0.16cm}\textbf{89.16} & 76.00 & 75.84 & 80.00 & 80.00 & 81.13 & 80.98 \\
     & SentiTPR & 91.75 & 91.73 & 92.85 & 92.85 & 85.64 & 85.53 & 78.00 & 77.96 & 84.00 & 84.00 & \textbf{82.97} & \hspace{0.17cm}\textbf{82.54} \\
 RF & None & 65.25 & 64.53 & 67.85 & 67.23 & 75.56 & 75.21 & 67.50 & 67.40 & 73.00 & 72.09 & 74.47 & 74.70 \\
     & Chi2* & 59.00 & 59.93 & 52.85 & 52.85 & 67.00 & 66.43 & 57.50 & 56.15 & 64.00 & 60.41 & 48.36 & 46.52 \\
     & Chi2** & 50.00 & 49.92 & 52.50 & 50.25 & 60.95 & 55.26 & 62.00 & 61.94 & 60.00 & 55.07 & 51.20 & 46.18 \\
     & PCA* & 61.25 & 60.66 & 56.78 & 56.53 & 71.28 & 71.28 & 60.00 & 60.00 & 67.00 & 66.90 & 61.46 & 60.98 \\
     & PCA** & 60.50 & 59.59 & 50.71 & 50.54 & 64.73 & 64.30 & 56.50 & 56.41 & 66.50 & 66.31 & 56.64 & 51.73 \\
     & GRP* & 52.75 & 51.97 & 51.42 & 50.38 & 58.43 & 58.41 & 57.00 & 56.84 & 63.00 & 62.76 & 55.93 & 54.74 \\
     & GRP** & 52.50 & 51.68 & 51.78 & 50.35 & 57.68 & 57.67 & 59.00 & 58.93 & 64.00 & 64.00 & 54.37 & 53.27 \\
     & KPCA* & 65.25 & 65.11 & 55.35 & 54.85 & 64.98 & 64.96 & 56.50 & 56.31 & 62.50 & 62.45 & 61.98 & 61.56 \\
     & KPCA** & 59.50 & 58.29 & 56.07 & 55.81 & 67.50  & 67.50 & 53.00 & 52.31 & 62.00 & 61.90 & 54.79 & 50.63 \\
     & SRP* & 53.00 & 50.72 & 48.21 & 46.92 & 75.06 & 74.92 & 62.00 & 61.98 & 68.00 & 67.95 & 70.40 & 70.20 \\
     & SRP** & 57.00 & 56.37 & 56.78 & 55.81 & 76.82 & 76.68 & 68.50 & 68.49 & 69.00 & 69.00 & 69.73 & 69.59 \\
	 & ISOMAP* & 53.00 & 50.72 & 48.21& 46.92 & 75.06 & 74.92 & 62.00 & 61.98 & 68.00 & 67.95 & 58.15 & 57.47 \\
	 & ISOMAP** & 57.00 & 56.37 & 56.78& 55.81 & 76.82 & 76.68 & 68.49 & 68.49 & 69.00 & 69.00 & 56.17 & 54.88 \\
     & SentiTPC & 80.50 & 80.40 & 78.92 & 78.72 & 83.12 & 83.10 & 75.00 & 74.34 & 80.00 & 79.70 & 76.31 & 76.11 \\
     & SentiTPR & 81.25 & 81.15 & 80.35 & 80.03 & 84.38 & 84.35 & 78.50 & 77.73 & 81.50 & 81.04 & 76.54 & 76.25 \\
	 
    \bottomrule
    \label{table:gene accuracy table}
  \end{tabular}
\end{table*}
\subsubsection{Average accuracy table for Movie domain datasets }
We first observe the average accuracy results belonging to the movie review domain dataset from Table \ref{table:averg accuarcy dataset1,2}. According to the linear SVM classifier, the Chi2* and Chi2** show the lowest average accuracy, while, on the other hand, SPR*, SPR**, ISOMAP*, and ISOMAP** show more effective performance. They improve the average accuracy by 18.65\%, 20.16\%, 19\%, and 20.16\% respectively.
The PCA*, PCA**, KPCA*, KPCA**, and baseline method (which is also called None) have superior performance to SPR*, SPR**, ISOMAP* and ISOMAP** reduction techniques. They enhance the average accuracy 5.7\%, 3.6\%, 5.4\%, 3.6\%, and 5.7\% respectively.
When we compared PCA*, PCA**, KPCA* and KPCA** with our proposed methods, SentiTPC and SentiTPR, we found out that SentiTPC and SentiTPR outperform these four techniques by an average accuracy of 14.32\%, 18.15\%, 14.32\%, and 18.15\% respectively. Thus, our proposed methods, SentiTPC, and SentiTPR show the highest overall performance across the movie domain datasets.

In terms of LR classifier which is similar to SVM classifier, Chi2* and Chi2** exhibit the lowest average accuracy whereas SPR*, SPR**, ISOMAP*, and ISOMAP** show much higher accuracy than Chi2* and Chi2**. Similarly, GPR* and GPR** show a slightly higher average accuracy than that of SPR*, SPR**, ISOMAP*, and ISOMAP** by an average increase of 0.47\% and 0.14\% respectively. PCA*, PCA** and baseline method i.e., None method, increase the average accuracy by a reasonable percentage of 4\% and 2.58\% than GPR* and GPR**. KPCA* and KPCA** outperform the PCA* and PCA** by an average accuracy of 1.76\%. Finally, our proposed methods (SentiTPC and SentiTPR) outperform KPCA* and KPCA** by 13.27\% and 17.86\% respectively. Thus, similarly, we can say that SentiTPC and SentiTPR show the highest accuracy among all reduction and non-reduction techniques. 
Likewise, for the NB classifier, our proposed methods again outperformed all the other reductions techniques. For the NB classifier, SentiTPC and SentiTPR show a higher average accuracy of 31.79\% and 37.16\% respectively, when compared with Chi2* and Chi2** and of 9.49\% and 14.47\% respectively when compared with baseline method. 
For random forest classifier, where Chi2*, Chi2**, SRP*, SRP**, ISOMAP*, and ISOMAP** show almost similar and low average accuracy across the movie domain datasets \ref{table:averg accuarcy dataset1,2}. Here, the baseline method shows a higher performance accuracy of 4.7\%, 7.2\%, 5.9\%, 12.9\%, 5.1\%, 9.5\% respectively, when compared with KPCA*, KPCA**, GPR*, GPR**, PCA* and PCA**. However, SentiTPC and SentiTPR outperform the baseline method by an average accuracy of 14.7\% and 15.8\% respectively. Thus, for random forest classifier, our proposed methods show the highest performance among all the other compared reduction techniques.\\

In our prior work \cite{ssci2019dr}, for movie domain datasets i.e., dataset 1 and dataset 2, the highest average accuracy achieved by TPC and TPR is 89.03\% and 79.98\% respectively which is archived by using LR classifier. However, when TPC and TPR are compared with our new proposed methods i.e., SentiTPC and SentiTPR, proposed methods is exhibited much higher performance and increased the average accuracy by 3.2\% and 16.8\% respectively.
\begin{table}
	\centering
	\caption{Average accuracy for movie domain datasets i.e., dataset 1 and dataset 2.}
	\begin{tabular}{p{1.7cm} p{1.2cm} p{1.2cm} p{1.2cm} p{1.2cm} }
		\toprule
		Reduction Techniques & SVM & LR & NB & RF \\    
		\midrule
		None & 77.35 & 77.00 & 77.83& 65.05 \\
		Chi2* & 52.925 & 55.66 & 55.53 & 55.92\\
		Chi2** & 53.55 & 52.76 & 51.14 & 51.25\\ 
		PCA* & 77.35 & 77.00 & N/A & 59.95 \\
		PCA** & 77.35 & 77.00 & N/A & 55.60 \\
		GRP* & 73.27 & 73.05 & N/A& 59.08 \\
		GRP** & 77.26 & 74.42 & N/A & 52.14 \\
		KPCA* & 77.35 & 78.96 & N/A & 60.30\\
		KPCA** & 77.35 & 78.96 & N/A & 57.78\\
		SRP* & 71.57 & 72.58 & N/A & 50.60\\
		SRP** & 73.71 & 74.25 & N/A & 56.89\\
		ISOMAP* & 71.92 & 72.58 & N/A & 50.60\\
		ISOMAP** & 73.71 & 74.28 & N/A & 56.89\\
		SentiTPC & 91.67 & 92.23 & 87.32 & 79.71\\
		SentiTPR & 95.50 & 96.82 & 92.30 & 80.80\\
	\bottomrule
	\label{table:averg accuarcy dataset1,2}
	\end{tabular}
\end{table}

\subsubsection{Average accuracy table for different review datasets }
In this part, we reported the experimental results on Arabic, Roman Urdu, and two English datasets. The average accuracy among these four datasets are listed in Table \ref{table:averg accuarcy dataset3,4,5,6}.
For the SVM classifier, SentiTPC and SentiTPR outperform all the reduction and non-reduction techniques. They show an increased average accuracy of 14.13\%, 10.93\%, 7.8\%, and 13.82\% when compared with ISOMAP*, ISOMAP**, SRP*, and SRP** respectively. In comparison with the baseline method (i.e., None), PCA*, PCA**, KPCA*, and KPCA**, the proposed method (SentiTPC) increased the accuracy by 11.12\%, 6.4\%, 7.87\%, 3.83\%, and 7.87\% respectively.

For LR classifier, PCA*, PCA**, KPCA*, KPCA**, and baseline method show higher average accuracy scores than all the other techniques except SentiTPC and SentiTPR. Which outperform these by an average accuracy of 8.8\%, 6.54\%, 8.9\%, 7.2\%, and 8.9\%. Similarly, for NB classifier, SentiTPC and SentiTPR outperform the baseline method, Chi2* and Chi2** by 4.8\%, 5.8\%, 20\% and 17.2\% respectively. 

Finally, for RF classifier, Chi2*, Chi2**, GRP*, and GRP** show the lowest accuracy among all the compared methods. Meanwhile, the PCA* and PCA** exhibit better performance when compared with these reduction techniques as mentioned above. Whereas ISOMAP*, ISOMAP**, SRP*, SRP**, and baseline method considerably outperform the PCA* and PCA**. However, the performance of SentiTPC and SentiTPR is better than ISOMAP*, ISOMAP**, SRP*, SRP**, and baseline method by an average accuracy score of 12.6\%, 12.8\%, 9.2\%, 9.7\%, 5.6\%, and 7.7\% respectively.

\begin{table}
	\centering
	\caption{Average accuracy table for different review datasets i.e., dataset 3, 4, 5 and 6.}
	\begin{tabular}{p{1.7cm} p{1.2cm} p{1.2cm} p{1.2cm} p{1.2cm} }
		\toprule
		Reduction Techniques & SVM & LR & NB & RF \\ 
		\midrule
		None & 67.16 & 74.47 & 76.79 & 72.63 \\
		Chi2* & 65.87 & 64.72 & 61.00 & 59.21\\
		Chi2** & 61.54 & 65.08 & 64.45 & 58.53\\ 
		PCA* & 71.85 & 76.10 & N/A & 64.93 \\
		PCA** & 67.16 & 74.45 & N/A & 61.09 \\
		GRP* & 64.20 & 64.39 & N/A & 58.59 \\
		GRP** & 68.11 & 72.26 & N/A & 58.76 \\
		KPCA* & 71.45 & 75.42 & N/A & 61.49\\
		KPCA** & 67.16 & 74.46 & N/A& 59.32\\
		SRP* & 64.46 & 67.31 & N/A& 68.86\\
		SRP** & 67.16 & 73.88 & N/A & 71.01\\
		ISOMAP* & 64.15 & 67.60 & N/A & 65.80\\
		ISOMAP** & 64.10 & 70.99 & N/A & 67.62\\
		SentiTPC & 78.28 & 82.64 & 81.57 & 78.60\\
		SentiTPR & 75.03 & 83.42 & 82.65 & 80.23\\
	\bottomrule
	\label{table:averg accuarcy dataset3,4,5,6}
	\end{tabular}
\end{table}

\subsection{Feature reduction with SentiTPC and SentiTPR}
\begin{figure}[h]
    \centering
    \includegraphics[width=0.48\textwidth]{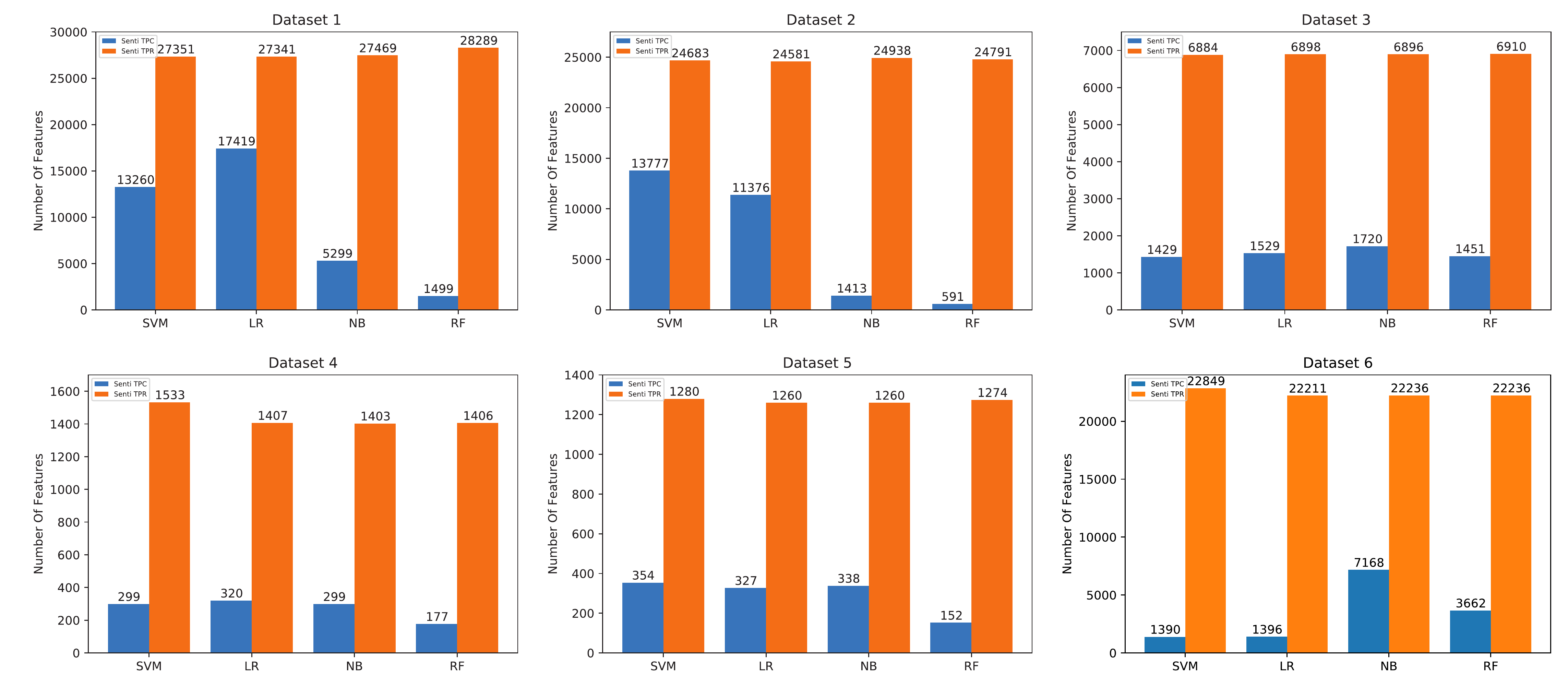}
    \caption{Selected number of features corresponding to SentiTPC and SentiTPR among various classifiers for dataset 1, 2, 3, 4, 5 and 6 respectively. }
    \label{fig:Number of features of Datasets}
\end{figure}

In this subsection, we try to elaborate Fig. \ref{fig:Number of features of Datasets}, which manifests the extent to which our proposed techniques (i.e., SentiTPC and SentiTPR) manage to reduce the feature size in order to maximize the accuracy score among different classifiers. The initial features of all the listed datasets can be seen in Table \ref{table:Chars of Data Sets}. For dataset 1, SVM with SentiTPC and SentiTPR reduce the features up to 68\% and 34\% respectively. While LR and NB along with SentiTPC, and SentiTPR reduce the feature sizes by approximately 58\%, 34\%, 87\%, and 34\% respectively. For RF classifier, SentiTPC and SentiTPR reduce the size of the initial set by around 96\% and 31\% respectively. Due to limited space, we have explicitly mentioned the feature reduction sizes only for dataset 1. Similarly, for other datasets, the extent of feature reduction sizes can very easily be analyzed from the Fig. \ref{fig:Number of features of Datasets}.

\subsection{Hyperparameters Sensitivity Study}
In this section, we study the sensitivity of the hyperparameters and how are the hyperparameters of our proposed methods influence the selected number of features and classification accuracy.
\begin{figure}[hbt!]
    \centering
    \includegraphics[width=0.5\textwidth]{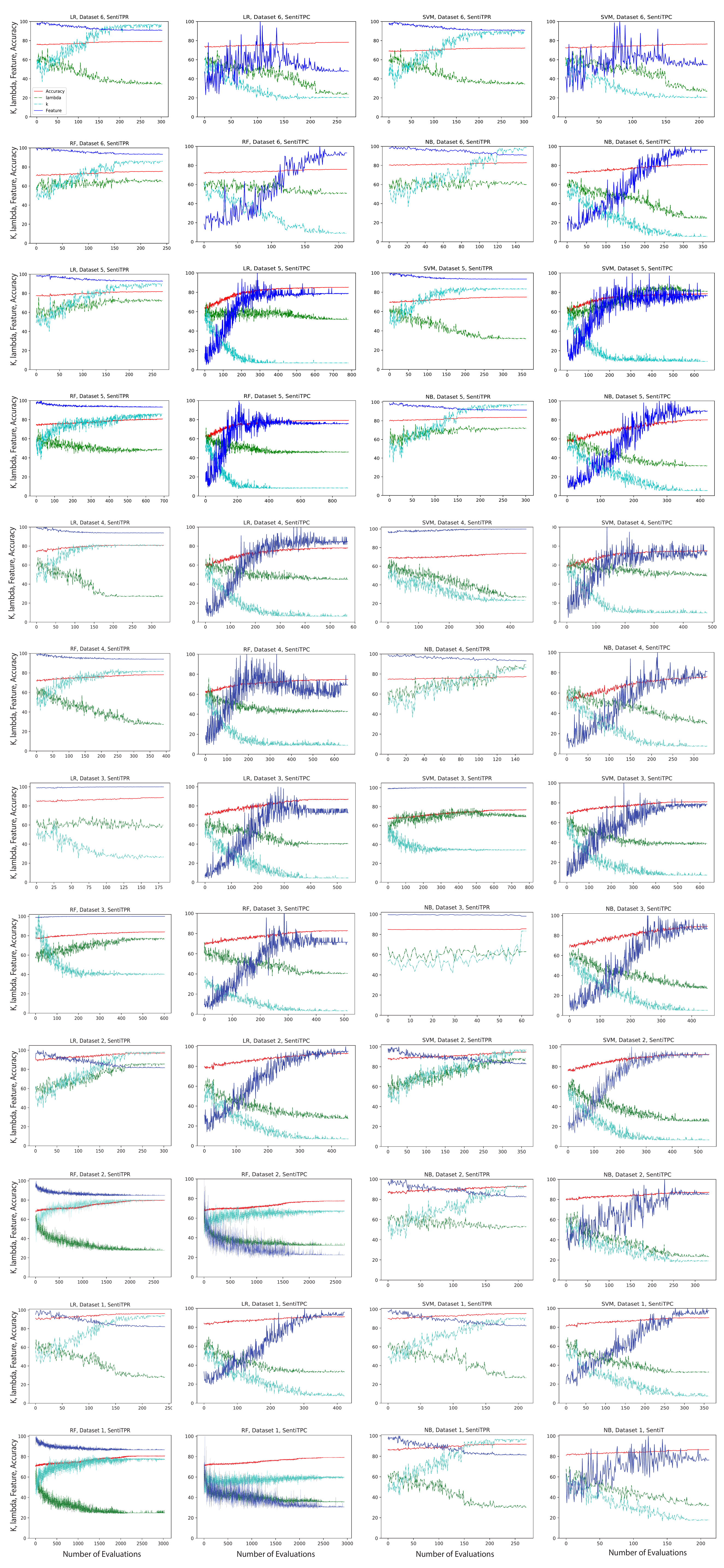}
    \caption{Demonstrate the convergence curves for 30 independent trials. It illustrates how the k and lambda values are evolved by evolutionary process and converged the k and lambda values in such a way that it selects the most prominent and separable features from the initial feature set in order to maximize the accuracy values.}
    \label{fig:convergence curves}
\end{figure}
\subsubsection{Hyperparameters setting of convergence curves}

Our proposed methods SentiTPC and SentiTPR, possess two hyperparameters (i.e., k and lambda). The range of the lambda parameter is same for both of the proposed methods, which is [0.1 -0.5]. However, hyperparameter k has different range of values, which are [1-30] and [1-50] for SentiTPC and SentiTPR respectively.
In the listed convergence curves, the x-axis is labeled as the number of evaluations. The total number of evaluations represent the number of times that DE evolves the hyper-parameters. 
The y-axis is labeled as k, lambda, accuracy, and feature values. As all these parameters incorporate different value ranges, so in order to make the convergence curve more consistent and coherent, we transform each parameter's values in the same range.
For this purpose, we multiply accuracy and lambda values by 100 and 200 respectively. Parameter k values are multiplied by 3.333 and 2 in SentiTPC and SentiTPR respectively. Similarly, to make the feature vector fall into the same range, we extract the maximum feature value from the feature vector and divide it by 100, which gives us a resultant value. Finally, each feature value is divided by this resultant value.

\subsubsection{Explanation of convergence curves}
The convergence curves in Fig. \ref{fig:convergence curves} reveal that how k and lambda values are being evolved by the differential evolution to extract the most prominent features from the feature set, which, as a result, maximizes the accuracy. As seen from the given curves, the relationship between k and lambda value, and accuracy and the number of features is highly nonlinear. That is the reason why the convergence curves are not very smooth for both of the methods. SentiTPC shows much higher fluctuations than SentiTPR because, at the start of the SentiTPC process, the k and lambda values are initialized in such a way that the number of feature values are extremely reduced. When these remarkably lower feature values are plotted into a much higher range, they produce very high fluctuations, and thus, the convergence curve is highly unsmooth.

At the start of the SentiTPC and SentiTPR methods, the k and lambda values are randomly initialized, and the features from the initial feature set are selected correspondingly. After that, the proposed methods compute the accuracy values based on the extracted features. However, with each passing evolution, the evolution process starts to learn and evolve the k and lambda value in such a way that it ensures to select the most prominent and separable features from the initial feature set. This, consequently, maximizes the accuracy.

The basic relation between the lambda and $Weight_{TPC} $, for SentiTPC method, is presented in the Equation \eqref{eq:EqsentiTPC}. This depicts that as lambda value decreases, the difference between the distribution distinction and total distribution information decreases, and as lambda value increases, the difference between distribution distinction and total distribution information increases. A similar phenomenon is being observed in SentiTPR methods, as shown in the Equation \eqref{eq: eqsentiTPR}. As the $Weight_{TPC} $, $Weight_{TPR}$ and k values have an explicit relationship with the selected number of features, when $Weight_{TPC} $, $Weight_{TPR}$ and k values increase, the selected number of features decrease and vice versa. But if the $Weight_{TPC} $ and $Weight_{TPR}$ are decreased, and the k-value is getting high, then ultimately, the number of features will be decreased and vice versa.\par
According to listed convergence curves for SentiTPR, at the start, the SentiTPR method, based on the k and lambda values, selects a bulk of irrelevant and noisy features from the initial feature set and produces lower accuracy. However, with the increasing number of evaluations, it increases the value of k and decreases the value of lambda, for most of the datasets except for dataset 3. This process of adjusting the k and lambda values accordingly puts a higher constraint on the proposed method to remove the insignificant and noisy features from the initial feature set. This property subsequently improves accuracy. On the other hand, for dataset 3, the lambda value increases, and so does the $Weight_{TPR}$ value, which, in turn, selects the most irrelevant features. Now, in order to exclude these irrelevant features, the k value decreases even further. As a result, a further increment in the accuracy value is achieved.\par
For SentiTPC convergence curve, at the start, k and lambda value are substantially high, which removes a large number of prominent features from the initial feature set. Thus, the selected number of features are considerably low at that point and, in turn, decreases the accuracy. So, to maximize the accuracy, the differential evolution starts evolving the k and lambda values from their higher values to the most suitable lower values and ultimately selects all the important features. The same phenomenon can be seen in the entire listed convergence curves of SentiTPC, except for RF with dataset 2 and dataset 3. For RF dataset 2 and dataset 3, the evolutionary process starts from higher values of lambda and k, and as a result, the $Weight_{TPC}$ values become extensively high, which ultimately selects the most insignificant features from the original feature set. Later on, the differential evolution method starts to converge k and lambda value, which reduces the number of features by removing noisy and irrelevant features in order to maximize the accuracy value.
\section{Conclusion}
In this paper, we introduced two dimensionality reduction techniques i.e., Sentiment Term Presence Count (SentiTPC) and Sentiment Term Presence Raito (SentiTPR) for sentiment classification which select or reject features based on their distribution distinction for SentiTPC and ratio of distribution distinction for SentiTPR. Moreover, these proposed techniques also consider the total distribution information of each feature in the initial feature set. In existing dimensionality reduction techniques, the number of components should be set manually, which can make the classification model uncertain. This may result in removing some prominent and most separable features hence reducing the performance of the classifiers.
Proposed SentiTPC and SentiTPR reject the features which have lower distribution and less distinctness. More significant and separable features will be kept, which helps the machine learning classifiers to make a clear decision boundary more conveniently, thus improving the classification performance. Experimental results manifest that SentiTPC and SentiTPR outperforms the competitive techniques on all of the datasets from different review domains in terms of classification accuracy and Avg.FM. For binary sentiment classification task, enhanced Avg.FM imply that our proposed methods are more generalized. Furthermore, hyperparameters i.e., k and lambda of SentiTPC and SentiTPR are evolved by the differential evolution. Besides, we employ Arabic and Urdu datasets which validate the fact that our proposed techniques are language independent.


\ifCLASSOPTIONcompsoc
  \section*{Acknowledgments}
\else
  \section*{Acknowledgment}
\fi
This work was supported by National Natural Science Foundation of China under Grant No. 61872419, No. 61573166, No. 61572230, No. 61873324, No. 81671785, No. 61672262, No. 61903156. Shandong Provincial Natural Science Foundation No. ZR2019MF040, No. ZR2018LF005. Shandong Provincial Key R\&D Program under Grant No. 2019GGX101041, No. 2018CXGC0706, No. 2017CXZC1206. Taishan Scholars Program of Shandong Province, China, under Grant No. tsqn201812077.
\ifCLASSOPTIONcaptionsoff
  \newpage
\fi



%


\bibliography{ref} 
\bibliographystyle{ieeetr}

%

\vspace{-2.05cm}
\begin{IEEEbiography}[
{
\includegraphics[width=1in,height=1.25in]{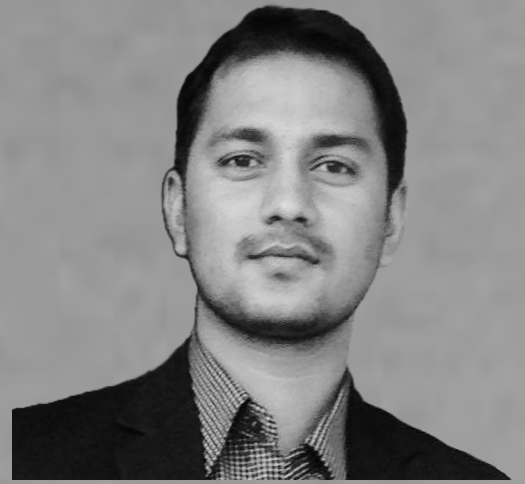}
}
]{Aftab Anjum} was born in 1991 in punjab Province of pakistan. He received his bachelor’s degree from University of Lahore, Pakistan, in 2017. Currently, he is pursuing his MS in Computer Science from School of Information Science and Engineering, University of Jinan, Jinan, 250022, China. His main areas of research interests include Machine Learning, Natural Language Processing, Sentiment Analysis, Evolutionary Computation and Mathematical Modeling.
\end{IEEEbiography}

\vskip -0.1\baselineskip plus -1fil

\begin{IEEEbiography}[
{
\includegraphics[width=1in,height=1.25in]{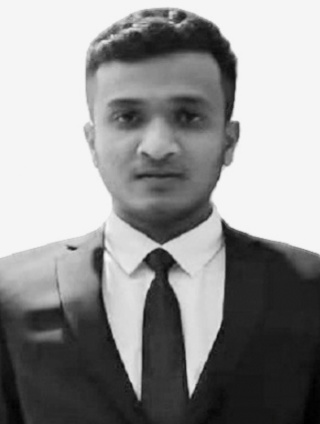}
}
]{Mazharul Islam}
graduated from International Islamic University Chittagong in May 2017 with a bachelor’s of science in Computer Science and Engineering. Currently, he is enrolled as a MS student at School of Information Science and Engineering, University of Jinan, Jinan, 250022, China. He has a background in Machine Learning, Natural Language Processing, and Data Mining and holds keen interests in the area of Sentiment Analysis for predicting the future from review or time series data. 
\end{IEEEbiography}
\vskip -0.1\baselineskip plus -1fil
\begin{IEEEbiography}
[{\includegraphics[width=1in,height=1.25in]{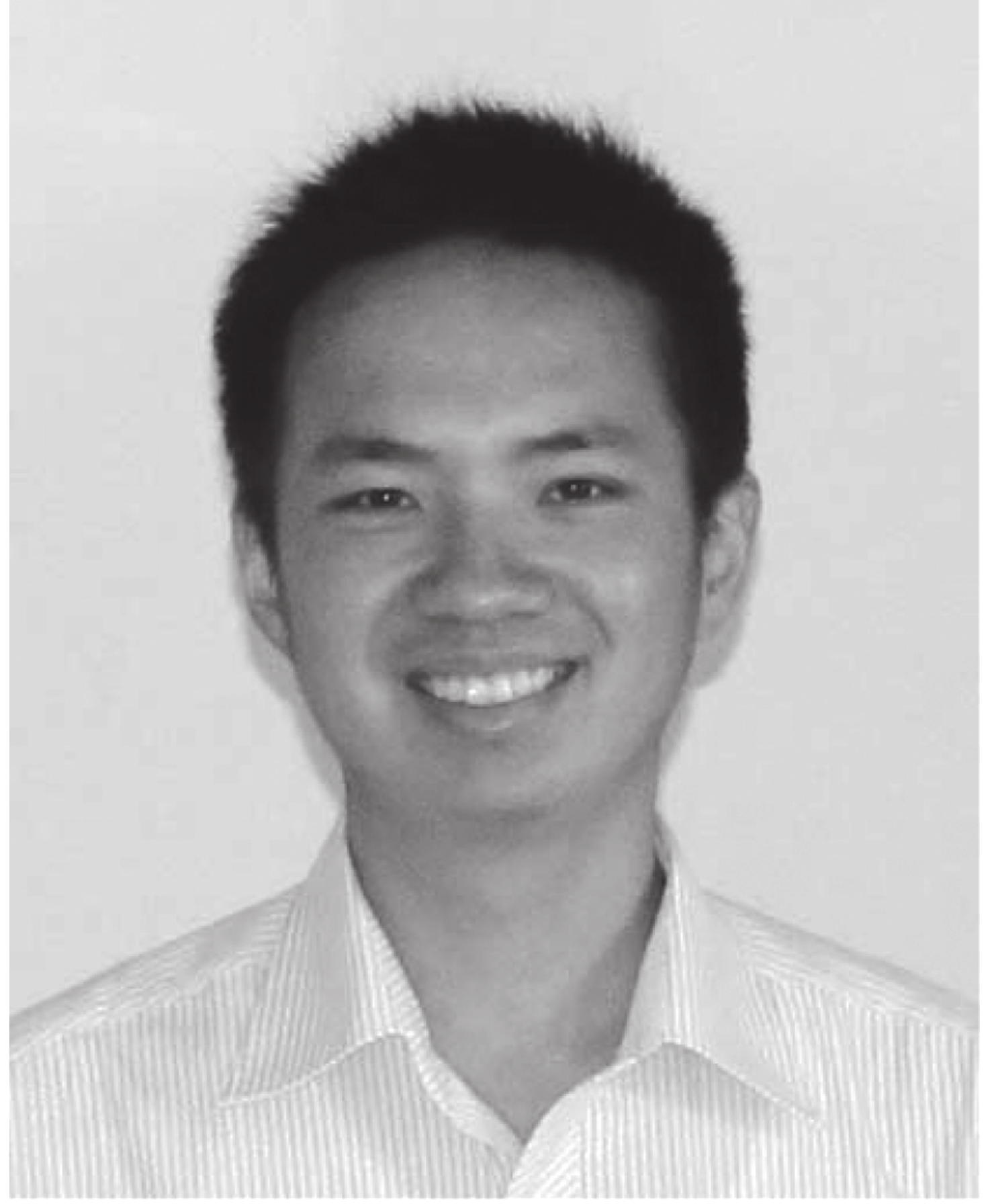}}]{Lin Wang} was born in 1983 in Shandong Province of China. In 2005 and 2008, He received his B.Sc. and Master degrees in Computer Science and Technology from the University of Jinan. In 2011, he received his Ph.D degree in Computer Science and Technology from the School of Computer science and technology, Shandong University, Jinan, China. Now, he is an associate professor in Shandong Provincial Key Laboratory of Network based Intelligent Computing, University of Jinan, Jinan, 250022, China. His research interests include Classification, Hybrid Computational Intelligence and Mathematical
Modeling.
\end{IEEEbiography}



\end{document}